\pdfoutput=1
\documentclass[10pt,journal,compsoc]{IEEEtran}

\usepackage{alltt}
\usepackage{multirow}
\usepackage{graphicx}
\usepackage{color}
\usepackage[labelfont=bf,textfont={bf}]{caption}
\usepackage{subcaption}
\usepackage{pifont}
\usepackage{boxedminipage}
\usepackage{flushend}
\usepackage{notoccite}
\usepackage{array}
\newcolumntype{C}[1]{>{\centering\let\newline\\\arraybackslash\hspace{0pt}}m{#1}}
\usepackage{booktabs}
\usepackage[framemethod=tikz]{mdframed}
\usepackage{lipsum}
\usetikzlibrary{shadows}
\usepackage{mathtools}
 \usepackage{paralist}
 \definecolor{light-gray}{gray}{0.80}
\usepackage{graphics}
\newmdenv[
tikzsetting= {fill=light-gray},
linewidth=1pt,
roundcorner=0pt, 
shadow=false
]{myshadowbox}
\usepackage{colortbl} 
\usepackage{subcaption}
\usepackage{blindtext, graphicx}
\usepackage{textcomp}
\usepackage{mathtools}
\usepackage{hyperref}
\usepackage{amsmath}
\usepackage{amsfonts}
\usepackage{caption}
\usepackage{color}
\usepackage[final]{listings}
\usepackage{graphicx} 
\usepackage{multirow}
\usepackage{balance}
\usepackage{picture}
\usepackage{relsize}
\usepackage{soul}
\usepackage{enumitem}
\setitemize{noitemsep,topsep=0pt,parsep=0pt,partopsep=0pt}
\DeclarePairedDelimiter\abs{\lvert}{\rvert}%
\usepackage{array}
\usepackage{makecell}

\definecolor{comment_color}{rgb}{0.5, 0, 1}

\newcommand{\fig}[1]{Figure~\ref{fig:#1}}
\newcommand{\bi}{\begin{itemize}}
\newcommand{\ei}{\end{itemize}}
\usepackage[
  pass,
]{geometry}
\usepackage{verbatim}
\usepackage{algorithm}
\usepackage{algorithmicx}
\usepackage{algpseudocode}
\usepackage[export]{adjustbox}
\usepackage{mathtools}
\renewcommand{\footnotesize}{\scriptsize}
\definecolor{lightgray}{gray}{0.8}
\definecolor{darkgray}{gray}{0.6}

\definecolor{Gray}{rgb}{0.88,1,1}
\definecolor{Gray}{gray}{0.85}
\definecolor{Blue}{RGB}{0,29,193}
\definecolor{MyDarkBlue}{rgb}{0,0.08,0.45} 
\usepackage{fancyvrb}
\fvset{%
fontsize=\small,
numbers=left
}

\newcommand{\quart}[3]{\begin{picture}(80,6)
{\color{black}\put(#3,3){\circle*{4}}\put(#1,3){\line(1,0){#2}}}\end{picture}}
\newcommand{\quartex}[3]{
\begin{picture}(13,6)
    {
     \color{black}
        \put(#3,3)
        {\circle*{4}}
        \put(#1,3)
        {\line(1,0){#2}}
    }
\end{picture}
}
\definecolor{lightgray}{gray}{0.7}
\tikzstyle{highlighter} = [
  lightgray,
  line width = \baselineskip,
]

\newcounter{highlight}[page]

\AtBeginShipout{\AtBeginShipoutUpperLeft{\ifthenelse{\value{highlight} > 0}{\tikz[remember picture, overlay]{\foreach \stroke in {1,...,\arabic{highlight}} \draw[highlighter] (begin highlight \stroke) -- (end highlight \stroke);}}{}}}

\newcommand{\squishlist}{
 \begin{list}{$\bullet$}
  { \setlength{\itemsep}{0pt}
     \setlength{\parsep}{3pt}
     \setlength{\topsep}{3pt}
     \setlength{\partopsep}{0pt}
     \setlength{\leftmargin}{1.5em}
     \setlength{\labelwidth}{1em}
     \setlength{\labelsep}{0.5em} } }

\newcommand{\squishlisttwo}{
 \begin{list}{$\bullet$}
  { \setlength{\itemsep}{0pt}
     \setlength{\parsep}{0pt}
    \setlength{\topsep}{0pt}
    \setlength{\partopsep}{0pt}
    \setlength{\leftmargin}{2em}
    \setlength{\labelwidth}{1.5em}
    \setlength{\labelsep}{0.5em} } }

\newcommand{\squishend}{
  \end{list}  }

\newcommand{\flash}{{\sc Flash}\xspace}
\usepackage{soul}
\usepackage{color}

\definecolor{awesome}{rgb}{1.0, 0.13, 0.32}

\usepackage{amsmath}
\definecolor{Gray}{gray}{0.95}
\definecolor{LightGray}{gray}{0.975}

\ifCLASSOPTIONcompsoc
  \usepackage[nocompress]{cite}
\else
  \usepackage{cite}
\fi

\begin{document}
%
\title{Finding Faster Configurations using {\sc FLASH}}

\author{Vivek Nair,~
        Zhe Yu,~
        Tim Menzies,~
        Norbert Siegmund,~
        and Sven Apel
\IEEEcompsocitemizethanks{\IEEEcompsocthanksitem V. Nair, Z. Yu and T. Menzies are with the Department
of Computer Science, North Carolina State University, Raleigh, USA.\protect\\
E-mail: \{vivekaxl, azhe9 and tim.menzies\}@gmail.com
\IEEEcompsocthanksitem N. Siegmund is with the Department of Computer Science and Engineering, Bauhaus-University Weimar, Germany.\protect\\
E-mail: norbert.siegmund@uni-weimar.de
\IEEEcompsocthanksitem S. Apel is with the Department of Informatics and Mathematics,
University of Passau, Innstr. 33, 94032 Passau, Germany.\protect\\
E-mail:apel@uni-passau.de}
\thanks{Manuscript received November XX, 2017.}}

 \markboth{IEEE TRANS SE. submitted Nov`17}%
{Shell \MakeLowercase{\textit{et al.}}: Bare Demo of IEEEtran.cls for Computer Society Journals}

\IEEEtitleabstractindextext{%
\begin{abstract}
Finding good configurations of a software system is often challenging since the number of configuration options can be large.
Software engineers often make poor choices about configuration or, even worse, they usually use a sub-optimal configuration in production, which leads to inadequate performance. To assist engineers in finding the better configuration, this article introduces   \flash, a sequential model-based method that sequentially explores the configuration space by reflecting on the configurations evaluated so far to determine the next best configuration to explore.
\flash scales up to software systems that defeat the prior state-of-the-art model-based methods in this area. \flash runs much faster than existing methods and can solve both single-objective and multi-objective optimization problems. The central insight of this article is to use the prior knowledge of the configuration space (gained from prior runs) to choose the next promising configuration. This strategy reduces the effort (i.e., number of measurements) required to find the better configuration. 
We evaluate \flash  using 30 scenarios based on 7 software systems to demonstrate that \flash saves effort in 100\% and 80\% of cases in single-objective and multi-objective problems respectively by up to several orders of magnitude compared to state-of-the-art techniques. 

\end{abstract}

}

\maketitle

\IEEEdisplaynontitleabstractindextext
\IEEEdisplaynontitleabstractindextext

\ifCLASSOPTIONcaptionsoff
  \newpage
\fi
\section{Introduction}\label{sec:intro}

\noindent
Most software systems available today are {\em configurable}; that is, they can be easily adjusted to achieve
a wide range of functional and  non-functional (e.g., energy or
performance) properties.  Once a configuration space
becomes large, it becomes difficult for humans  to keep track of the interactions between the configuration
options. Section 2 of this article offers more details on many of the problems seen with software configuration. In summary:
\begin{itemize}[leftmargin=*]
\item Many software systems have poorly chosen defaults~\cite{van2017automatic,herodotou2011starfish}. Hence, it is useful to seek better configurations.
\item Understanding the configuration space of software systems with large configuration spaces is challenging~\cite{xu2015hey}.
\item Exploring more than just a handful of configurations is usually infeasible due to long benchmarking time~\cite{zhu2017optimized}.
\end{itemize}
This article describes \flash,  a novel way to find better configurations for a software system (for a given workload). 
\flash is
a sequential model-based method (SMBO)~\cite{alipourfard2017cherrypick, snoek2012practical, brochu2010tutorial}
that reflects
on the evidence (configurations) retrieved at some point to select the estimated best
configuration to measure next.
This way, \flash  uses fewer evaluations to find better configurations compared to more expensive prior work~\cite{guo2013variability, sarkar2015cost, nair17, nair2017using, guo2017data}.

Prior work in this area primarily used two strategies. Firstly, researchers used machine learning to model the configuration space. The model is built sequentially, where new configurations are sampled randomly, and the quality or accuracy of the model is measured using a holdout set. The size of the holdout set in some cases could be up to 20\% of the configuration space~\cite{nair2017using}  and needs to be evaluated (i.e., measured) before even the model is fully built. This strategy makes these methods not suitable in a practical setting since the generated holdout set can be (very) expensive. Secondly, the sequential model-based techniques used in prior work relied on Gaussian Process Models (GPM) to reflect on the configurations explored (or evaluated) so far~\cite{zuluaga2016varepsilon}. However, GPMs do not scale well for software systems with more than a dozen configuration options~\cite{wang2016bayesian}. 

The key idea of \flash is to build a performance model that is just accurate enough for differentiating better configurations from the rest of the configuration space. Tolerating the inaccuracy of the model is useful to reduce the cost (measured in terms of the number of configurations evaluated) and the time required to find the better configuration. To increase the scalability of methods using GPM, \flash replaces the GPMs with a fast and scalable decision tree learner.

 \begin{table*}
\centering
    \renewcommand{\arraystretch}{1} 
\resizebox{\linewidth}{!}{
    \begin{tabular}{p{1.2cm}p{1.4cm}p{1.65cm}rp{4.85cm}p{6.85cm}crp{0.85cm}}
\toprule
\textbf{Family} & \textbf{Software Systems} & \textbf{Objectives} & \textbf{\begin{tabular}[c]{@{}r@{}}\#Config\\ Options\end{tabular}} & \textbf{Configuration Options} & \textbf{Description} & \textbf{Abbr} & \textbf{\begin{tabular}[c]{@{}r@{}}\# Configu-\\ rations\end{tabular}} & \textbf{Prev Used} \\ \midrule
\multirow{18}{*}{\parbox{1\linewidth}{\vspace{3cm}\textbf{\begin{tabular}[c]{@{}l@{}}Stream\\ Processing \\ Systems\end{tabular}}}} & \multirow{2}{*}{wc-c1-3d} & Throughput & \multirow{2}{*}{3} & \multirow{2}{*}{max spout, spliters, counters} & \multirow{2}{*}{\begin{tabular}[c]{@{}l@{}}Word Count is executed by varying 3 configurations of\\ Apache Storm on cluster C1\end{tabular}} & SS-A1 & \multirow{2}{*}{1343} & \multirow{18}{*}{\parbox{1\linewidth}{\vspace{3.7cm}\cite{jamshidi2016uncertainty}}} \\ \cmidrule(lr){3-3} \cmidrule(lr){7-7}
 &  & Latency &  &  &  & SS-A2 &  &  \\ \cmidrule(lr){2-8}
 & \multirow{2}{*}{wc-c3-3d} & Throughput & \multirow{2}{*}{3} & \multirow{2}{*}{max spout, spliters, counters} & \multirow{2}{*}{\begin{tabular}[c]{@{}l@{}}Word Count is executed by varying 3 configurations of\\ Apache Storm on cluster C3\end{tabular}} & SS-C1 & \multirow{2}{*}{1512} &  \\ \cmidrule(lr){3-3} \cmidrule(lr){7-7}
 &  & Latency &  &  &  & SS-C2 &  &  \\ \cmidrule(lr){2-8}
 & \multirow{2}{*}{wc+wc-c4-3d} & Throughput & \multirow{2}{*}{3} & \multirow{2}{*}{max spout, spliters, counters} & \multirow{2}{*}{\begin{tabular}[c]{@{}l@{}}Word Count is executed, collocated with Word Count\\ task, by varying 3 configurations of Apache Storm on\\ cluster C3\end{tabular}} & SS-D1 & \multirow{2}{*}{195} &  \\ \cmidrule(lr){3-3} \cmidrule(lr){7-7}
 &  & Latency &  &  &  & SS-D2 &  &  \\ \cmidrule(lr){2-8}
 & \multirow{2}{*}{wc-c4-3d} & Throughput & \multirow{2}{*}{3} & \multirow{2}{*}{max spout, spliters, counters} & \multirow{2}{*}{\begin{tabular}[c]{@{}l@{}}Word Count is executed by varying 3 configurations of\\ Apache Storm on cluster C4\end{tabular}} & SS-E1 & \multirow{2}{*}{756} &  \\ \cmidrule(lr){3-3} \cmidrule(lr){7-7}
 &  & Latency &  &  &  & SS-E2 &  &  \\ \cmidrule(lr){2-8}
 & \multirow{2}{*}{wc+rs-c4-3d} & Throughput & \multirow{2}{*}{3} & \multirow{2}{*}{max spout, spliters, counters} & \multirow{2}{*}{\begin{tabular}[c]{@{}l@{}}Word Count is executed, collocated with Rolling Sort\\ task, by varying 3 configurations of Apache Storm on\\ cluster C4\end{tabular}} & SS-F1 & \multirow{2}{*}{196} &  \\ \cmidrule(lr){3-3} \cmidrule(lr){7-7}
 &  & Latency &  &  &  & SS-F2 &  &  \\ \cmidrule(lr){2-8}
 & \multirow{2}{*}{wc+sol-c4-3d} & Throughput & \multirow{2}{*}{3} & \multirow{2}{*}{max spout, spliters, counters} & \multirow{2}{*}{\begin{tabular}[c]{@{}l@{}}Word Count is executed, collocated with SOL task, by\\ varying 3 configurations of Apache Storm on cluster\\ C3\end{tabular}} & SS-G1 & \multirow{2}{*}{195} &  \\ \cmidrule(lr){3-3} \cmidrule(lr){7-7}
 &  & Latency &  &  &  & SS-G2 &  &  \\ \cmidrule(lr){2-8}
 & \multirow{2}{*}{wc-5d-c5} & Throughput & \multirow{2}{*}{5} & \multirow{2}{*}{\begin{tabular}[c]{@{}l@{}}spouts, splitters, counters,buffer-size,\\ heap\end{tabular}} & \multirow{2}{*}{\begin{tabular}[c]{@{}l@{}}Word Count is executed by varying 5 configurations of\\ Apache Storm on cluster C3\end{tabular}} & SS-I1 & \multirow{2}{*}{1080} &  \\ \cmidrule(lr){3-3} \cmidrule(lr){7-7}
 &  & Latency &  &  &  & SS-I2 &  &  \\ \cmidrule(lr){2-8}
 & \multirow{2}{*}{rs-6d-c3} & Throughput & \multirow{2}{*}{6} & \multirow{2}{*}{\begin{tabular}[c]{@{}l@{}}spouts, max spout, sorters, emitfreq,\\ chunksize, message size\end{tabular}} & \multirow{2}{*}{\begin{tabular}[c]{@{}l@{}}Rolling Sort is executed by varying 6 configurations of\\ Apache Storm on cluster C3\end{tabular}} & SS-J1 & \multirow{2}{*}{3839} &  \\ \cmidrule(lr){3-3} \cmidrule(lr){7-7}
 &  & Latency &  &  &  & SS-J2 &  &  \\ \cmidrule(lr){2-8}
 & \multirow{2}{*}{wc-6d-c1} & Throughput & \multirow{2}{*}{6} & \multirow{2}{*}{\begin{tabular}[c]{@{}l@{}}spouts, max spout, sorters,emitfreq,\\ chunksize,,message size\end{tabular}} & \multirow{2}{*}{\begin{tabular}[c]{@{}l@{}}Word Count is executed by varying 6 configurations of\\ Apache Storm on cluster C1\end{tabular}} & SS-K1 & \multirow{2}{*}{2880} &  \\ \cmidrule(lr){3-3} \cmidrule(lr){7-7}
 &  & Latency &  &  &  & SS-K2 &  &  \\ \midrule
\multirow{4}{*}{\textbf{FPGA}} & \multirow{2}{*}{sort-256} & Area & \multirow{2}{*}{3} & \multirow{2}{*}{Not specified} & \multirow{2}{*}{\begin{tabular}[c]{@{}l@{}}The design space consists of 206 different hardware\\ implementations of a sorting network for 256 inputs\end{tabular}} & SS-B1 & \multirow{2}{*}{206} & \multirow{6}{*}{\parbox{1\linewidth}{\vspace{1.2cm}\cite{zuluaga2016varepsilon}}} \\ \cmidrule(lr){3-3}
 &  & Throughput &  &  &  & SS-B2 &  &  \\ \cmidrule(lr){2-8}
 & \multirow{2}{*}{noc-CM-log} & Energy & \multirow{2}{*}{4} & \multirow{2}{*}{Not specified} & \multirow{2}{*}{\begin{tabular}[c]{@{}l@{}}The design space consists of 259 different implementations\\of a tree-based network-on-chip,targeting application specific\\circuits (ASICs) and,multi-processor system-on-chip designs\end{tabular}} & SS-H1 & \multirow{2}{*}{259} &  \\ \cmidrule(lr){3-3}
 &  & Runtime &  &  &  & SS-H2 &  &  \\ \cmidrule(r){1-8}
\multirow{2}{*}{\textbf{Compiler}} & \multirow{2}{*}{llvm} & Performance & \multirow{2}{*}{11} & \multirow{2}{*}{\begin{tabular}[c]{@{}l@{}}time passes, gvn, instcombine,inline,\\ ... , ipsccp, iv users, licm\end{tabular}} & \multirow{2}{*}{\begin{tabular}[c]{@{}l@{}}The design space consists of 1023 different compiler settings\\for the LLVM compiler framework. Each setting is specified by\\d = 11 binary flags.\end{tabular}} & SS-L1 & \multirow{2}{*}{1023} &  \\ \cmidrule(lr){3-3} \cmidrule(lr){7-7}
 &  & Memory Footprint &  &  &  & SS-L2 &  &  \\ \midrule
\multirow{2}{*}{\textbf{\begin{tabular}[c]{@{}l@{}}Mesh\\ Solver\end{tabular}}} & \multirow{2}{*}{Trimesh} & \# Iteration & \multirow{2}{*}{13} & \multirow{2}{*}{\begin{tabular}[c]{@{}l@{}}F, smoother, colorGS,relaxParameter, V, Jac-\\obi, line, zebraLine, cycle, alpha, beta,preSmo-\\othing, postSmoothing\end{tabular}} & \multirow{2}{*}{\begin{tabular}[c]{@{}l@{}}Configuration space of Trimesh, a library to\\ manipulate triangle meshes\end{tabular}} & SS-M1 & \multirow{2}{*}{239,260} & \multirow{6}{*}{\parbox{1\linewidth}{\vspace{1.5cm}\cite{siegmund2012predicting}}} \\ \cmidrule(lr){3-3}
 &  & Time to Solutions &  &  &  & SS-M2 &  &  \\ \cmidrule(r){1-8}
\multirow{2}{*}{\textbf{\begin{tabular}[c]{@{}l@{}}Video\\ Encoder\end{tabular}}} & \multirow{2}{*}{X264-DB} & PSNR & \multirow{2}{*}{17} & \multirow{2}{*}{\begin{tabular}[c]{@{}l@{}}no mbtree, no asm, no cabac, no\\ scenecut,..., keyint, crf, scenecut,\\ seek, ipratio\end{tabular}} & \multirow{2}{*}{Configuration space of X-264 a video encoder} & SS-N1 & \multirow{2}{*}{53,662} &  \\ \cmidrule(lr){3-3}
 &  & Energy &  &  &  & SS-N2 &  &  \\ \cmidrule(r){1-8}
\multirow{2}{*}{\textbf{\begin{tabular}[c]{@{}l@{}}Seismic\\ Analysis\\ Code\\~~~\end{tabular}}} & \multirow{2}{*}{SaC} & Compile-Exit & \multirow{2}{*}{59} & \multirow{2}{*}{\begin{tabular}[c]{@{}l@{}}extrema, enabledOptimizations,\\ disabledOptimizations,ls, dcr, cf, lir, inl,\\ lur, wlur, ...maxae, initmheap, initwheap\end{tabular}} & \multirow{2}{*}{Configuration space of SaC} & SS-O1 & \multirow{2}{*}{65,424} &  \\ \cmidrule(lr){3-3}
 &  & Compile-Read &  &  &  & SS-O2 &  &  \\ \bottomrule
\end{tabular}
}
\caption{
Configuration problems explored in this article.   The abbreviations of the systems (Abbr) are sorted in the order of the number of configuration options of the system. The column \#Config Options represent the number of configuration options of the software system and \#Configurations represents the total number of configurations of the system. See \url{http://tiny.cc/flash_systems/} for more details.
}
\label{fig:software_systems}
\end{table*}

\noindent The novel contributions of the article are:
\begin{itemize}[leftmargin=*]
    \item  We show that \flash can  solve single-objective performance configuration optimization problems using
    an order of magnitude fewer measurements than the
    state-of-the-art (Section~\ref{sec:so_results}). This is a critical feature 
    because, as discussed in Section~\ref{sec:perf_opt}, it can be very slow
    to sample multiple properties of modern software systems, when each
    such sample requires (say) to compile and benchmark the corresponding software system.
    \item We empower \flash to multi-objective performance configuration optimization problems.
    \item We show that \flash   overcomes the shortcomings of prior work and achieves similar performance and scalability, with greatly reduced runtimes (Section~\ref{sec:mo_results}).
    \item Background material, a replication package, all measurement data, and
the open-source version of \flash
are available at supplementary website (\url{http://tiny.cc/flashrepo/}).
\end{itemize}

The rest of the article is structured as follows: Section~\ref{sec:perf_opt} motivates this work. Section 3 describes the problem formulation and the theory behind SMBO. Section 4 describes prior work in software performance configuration optimization, followed by the core algorithm of \flash in Section 5. In Section 6, we present our research questions along with experimental settings used to answer them. Prior work in this area addresses a single-objective problem with the only exception of ePAL~\cite{zuluaga2016varepsilon}. Hence, we evaluate \flash separately for single-objective and multi-objective performance configuration optimization problems. In Section 7, we apply \flash on single-objective performance configuration optimization and multi-objective performance configuration optimization. 
The article ends with a discussion on various aspects of \flash, and finally, we conclude along with a discussion of future work.

\section{Performance configuration optimization for Software} \label{sec:perf_opt}

 This section  motivates our research by 
 reviewing the numerous problems associated with 
  software configuration.

Many researchers report that modern software systems come with \textcolor{black}{{\em a daunting number of configuration options}}. 
For example, the number of configuration options in Apache (a popular web server) increased from 150 to more than 550 configuration options within 16 years~\cite{xu2015hey}. 
Van Aken et al.~\cite{van2017automatic} also reports a similar trend. They indicate that, in over 15 years, the number of configuration options of {\sc Postgres} and {\sc MySQL} increased by a factor of three and six, respectively.
This is troubling since
Xu et al.~\cite{xu2015hey} report that developers tend to ignore over 80\% of configuration options, which leaves considerable optimization potential untapped and induces major economic cost~\cite{xu2015hey}.\footnote{The size of the configuration space increases exponentially  with the number of configuration options.} 
For illustration, \fig{software_systems} offer examples of the kinds of configuration options seen in software systems.

Another problem  with configurable
systems is the issue of  \textcolor{black}{{\em poorly chosen default configurations}}.
Often, it is assumed that software architects provide useful default configurations of their systems.  This assumption can be very
misleading.   Van Aken et al. report that the default MySQL configurations in 2016 assume that it will be installed on a machine that has  160MB of RAM (which, at that
time, was incorrect by, at least, an order of magnitude)~\cite{van2017automatic}. Herodotou et al.~\cite{herodotou2011starfish} show how standard settings for text mining
applications in Hadoop result in worst-case execution times.
In the same vein,  
Jamshidi et al.~\cite{jamshidi2016uncertainty} reports for
 text mining applications on Apache Storm, the throughput achieved using the worst configuration is 
{\em 480 times slower} than the throughput achieved by the best configuration.

Yet another  problem
is that 
\textcolor{black}{{\em  exploring benchmark sets for different configurations is very slow}}. 
Wang et al.~\cite{wang2013searching} comments on the problems of evolving a test suite for software if every candidate solution requires a time-consuming execution of the entire system: such test suite generation can take weeks of execution time.
Zuluaga et al.~\cite{zuluaga2013active} report on the cost of analysis for software/hardware co-design: ``synthesis of only one design can take hours or even days''.

The challenges of having numerous configuration options are just \textcolor{black}{{\em not limited to software systems}}. The problem to find an good set of configuration options is pervasive and faced in numerous other sub-domains of computer science and beyond. In software engineering, software product lines--where the objective is to find a product which (say) reduces cost and defects~\cite{chen2016sampling, henard2015combining}---have been widely studied.  
The problem of configuration optimization is present in domains, such as machine learning, cloud computing, and software security.

The area of \textcolor{black}{{\em hyper-parameter optimization}} (a.k.a.  parameter tuning) is very similar to the performance configuration optimization problem studied in this article. Instead of optimizing the performance of a software system, the hyper-parameter method tries to optimize the performance of a machine learner. Hyper-parameter optimization is an active area of research in various flavors of machine learning. For example, Bergstra and Bengiol~\cite{bergstra2013making} showed how random search could be used for hyper-parameter optimization of high dimensional spaces.
Recently, there has been much interest in 
hyper-parameter optimization applied to the area of software analytics~\cite{fu2016tuning, fufse17, fu2016differential, tantithamthavorn2016automated, agrawal2016wrong}.

Another area of application for performance configuration optimization is \textcolor{black}{{\em cloud computing}}.  
With the advent of big data, long-running analytics jobs are commonplace. Since different analytic jobs have diverse behaviors and resource requirements, choosing the correct virtual machine type
in a  cloud environment has become critical. This problem has received considerable interest, and we argue, this is another useful application of performance configuration optimization --- that is, optimize the performance of a system while minimizing cost~\cite{alipourfard2017cherrypick, venkataraman2016ernest, yadwadkar2017selecting, Zhu:2017:BTP:3127479.3128605, dalibard2017boat}.

As a sideeffect of the wide-spread adoption of cloud computing, the \textcolor{black}{{\em security}} of the instances or virtual machines (VMs) has become a daunting task. In particular, optimized security settings are not identical in every setup. They depend on characteristics of the setup, on the ways an application is used or on other applications running on the same system. The problem of finding security setting for a VM is similar to performance configuration optimization~\cite{biedermann2014hot, biedermann2014leveraging, drabik2003method, security1, security2}. 
Among numerous other problems which are similar to performance configuration optimization, the problem of how to maximize conversions on landing pages or click-through rates on search-engine result pages~\cite{hill2017efficient, wang2016beyond, zhu2017optimized} has gathered interest.

The rest of this article discusses how \flash addresses configuration problems (using
the case studies of \fig{software_systems}). 

\begin{figure*}[!htb]
    \minipage{0.32\textwidth}
      \includegraphics[width=\linewidth, height=1in]{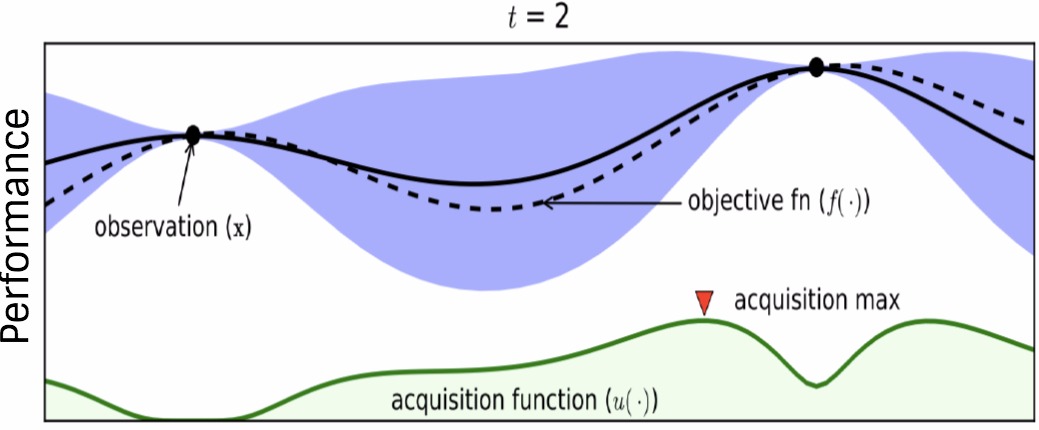}
    \endminipage\hfill
    \minipage{0.32\textwidth}
      \includegraphics[width=\linewidth, height=1in]{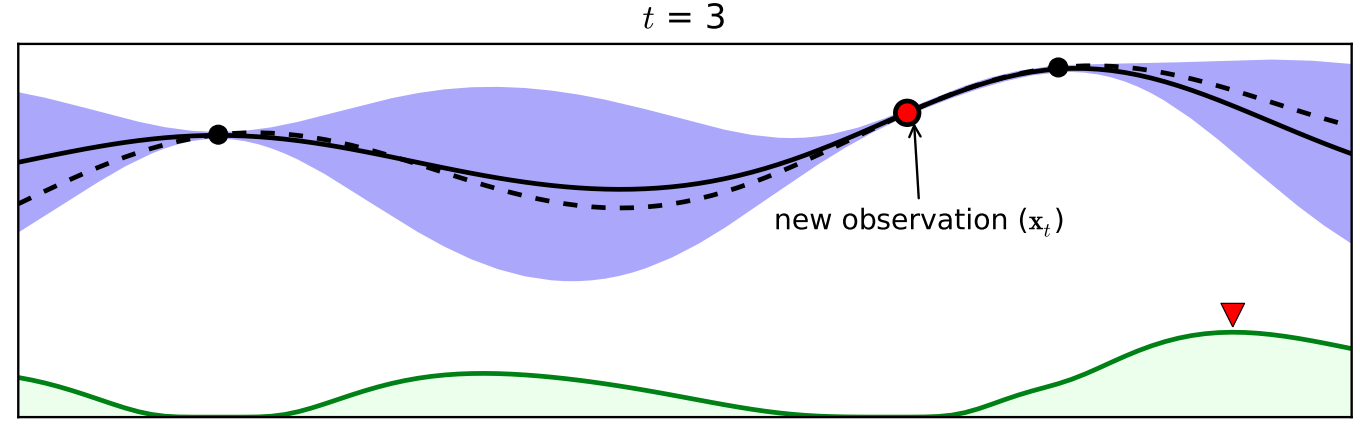}
    \endminipage\hfill
    \minipage{0.32\textwidth}%
      \includegraphics[width=\linewidth, height=1in]{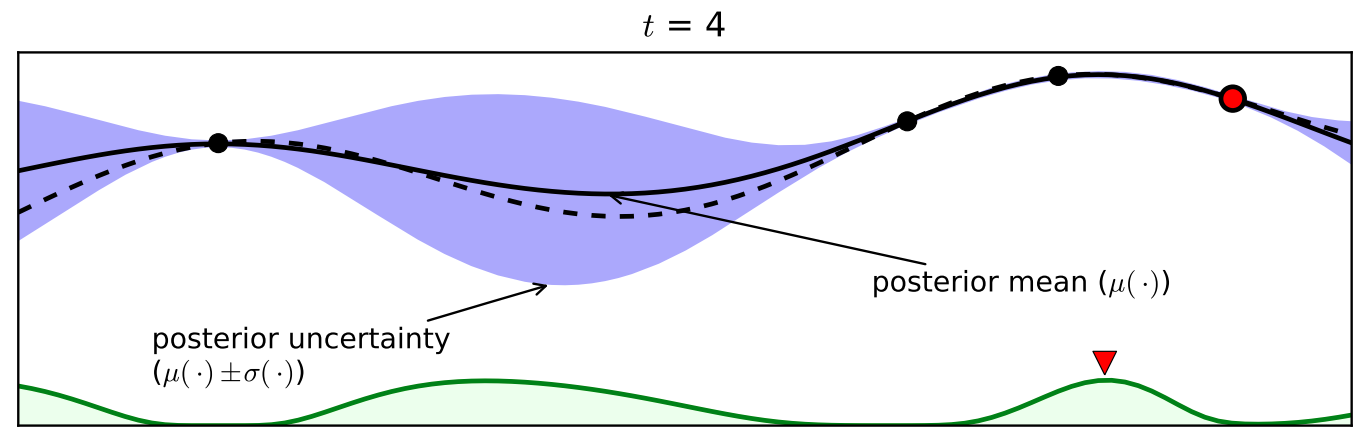}
    \endminipage
\caption{An example of Sequential Model-based method's working process from \cite{brochu2010tutorial}. The figures show a Gaussian process model (GPM) of an objective function over
four iterations of sampled values.
Green shaded plots represent acquisition function. The value of the acquisition function is high where the GPM predicts larger objective and where the prediction uncertainty (confidence) is high such points (configurations in our case) is sampled first. Note that the area on the far left is never sampled even when it has high uncertainty (low confidence) associated.} 
\label{fig:bayesian_optimazation}
\end{figure*}

\section{Theory}
The following theoretical notes define the framework used throughout the article.

\subsection{What are Configurable Software Systems?}\label{sec:problem_formal}

A configurable software system has a set $X$ of configurations $x \in X$. 
Let $x_i$ represent the \textit{ith} configuration of a software system. $x_{i,j}$ represent the \textit{jth} configuration option of the configuration $x_i$. In general, $x_{i,j}$ indicates either an (i) integer variable or a (ii) Boolean variable. The configuration space ($X$) represents all the valid configurations of a software system. The configurations are also referred to as \textit{independent variables}.
Each configuration ($x_i$), where $1\le i\le |X|$, has one (single-objective) or many (multi-objective) corresponding performance measures $y_{i,k} \in Y$, where $y_{i,k}$  indicates the $1\le kth\le m$ objective  associated with a configuration $x_i$. The performance measure is also referred to as \textit{dependent variable}. For multi-objective problems, there are multiple dependent variables. 
We denote the performance measures ($y\in Y$) associated with a given configuration by $(y_{i,1},..y_{i,m})=f(x_i)$, in multi-objective setting $y_i$ is a vector, where: $f: X\mapsto Y $ is a function which maps $X \in R^n$ to $Y \in R^m$. In a practical setting, whenever we try to obtain the performance measure corresponding to a certain configuration, it requires actually executing a benchmark run with that configuration. In our setting, evaluation of a configuration (or using $f$) is expensive and is referred to as a measurement. The cost or measurement is defined as the number of times $f$ is used to map a configuration $x_i \in X$ to $Y$. In our setting, the cost of an optimization technique is the total number of measurements required to find the better solution.

In the following, we will explore two different kinds of configuration optimization: {\em single-objective}
and {\em multiple objective}.
In single-objective performance configuration optimization,
we consider the problem of finding a good configuration ($x^*$) such that $f(x^*)$ is less than other configurations in $X$. Our objective is to find $x^*$ while minimizing the number of measurements.
\begin{equation}
    f(x^*) \le f(x),~~~ \forall x \in {X\setminus x^*}
\end{equation}
 
\noindent That is, our goal is to find the better configuration of a system with least cost or measurements as possible when compared to prior work.

In multi-objective performance configuration optimization,
we consider the problem of finding a  configuration ($x^*$) that is better than other configurations in the configuration space of $X$ while minimizing the number of measurements. Unlike, the single-objective configuration optimization problem, where one solution can be the best (optimal) solution (except multiple configurations have the same performance measure), in multi-objective configuration optimization there may be no best solution (best in all objectives). Rather there may be a set of solutions that are equivalent to each other. Hence,
to declare that one solution is better than another, all objectives must be polled separately. Given two vectors of configurations $x_1,x_2$ with associated objectives $y_{1}, y_{2}$,
then $x_1$ is binary dominant ($\succ$) over $x_2$ when:
\begin{equation}\label{eq:bdom}
    \begin{gathered}
            y_{1,p} \le y_{2,p} \forall p \in \{1, 2,..., m\}~~\text{and}\\
            y_{1,q} < y_{2, q}~~\text{for at least one index}~~q \in \{1, 2,...,m\}
    \end{gathered}
\end{equation}
where \textit{$y\in Y$} are the performance measures. We refer to binary dominant configurations as better configurations. For the multi-objective configuration optimization problem, our goal is to find a set of  better configurations of a given software system using fewer measurements compared to prior work.


\subsection{Sequential Model-based Optimization}
Sequential Model-based Optimization (SMBO) is a useful strategy to find extremes
of an unknown objective (or performance) function which is expensive (both
in terms of cost and time) to evaluate. In literature, a certain variant of SMBO is also called \textcolor{black}{Bayesian optimization}. SMBO
is efficient because of its ability to incorporate prior belief as already measured solutions (or configurations),
to help direct further sampling. Here, the prior represents
the already known areas of the search (or performance
optimization) problem. The prior can be used to estimate the
rest of the points (or unevaluated configurations). Once we have evaluated one (or many) points based on the
prior, we can define the posterior. The posterior captures
our updated belief in the objective function. This step is
performed by using a machine learning model, also called surrogate model.

\noindent The concept of SMBO is simple stated:
\bi
\item
Given what we know about the problem...
\item
... what should we do next?
\ei
The ``given what we know about the problem'' part is achieved by using a \textit{machine learning model} whereas ``what should we do next'' is performed by an \textit{acquisition function}. Such acquisition function automatically adjusts the exploration (``should we sample in uncertain parts of the search space) and exploitation (``should we stick to what is already known'') behavior of the method. 

This can also be explained as follows. Firstly, few points (or configurations) are (say) randomly selected and measured. These points along with their performance measurements are used to build a model (prior). Secondly, this model is then used to estimate or predict the performance measurements of other unevaluated points (or configurations).
This can be used by an acquisition function to select the configurations to measure next. This process continues till a predefined stopping criterion (budget) is reached.

Much of the prior  research in configuration optimization of software systems can
be characterized as an exploration of 
different acquisition functions.
These acquisition function (or sampling heuristics) were used to satisfy two requirements: (i) use a `reasonable' number of configurations (along with corresponding measurements) and (ii) the selected configurations should incorporate the relevant interactions---how different configuration options influence the performance measure~\cite{Siegmund2015}.
In a nutshell, the intuition behind such functions is that it is not necessary to try all configuration options---for pragmatic reasons. Rather, it is only necessary to try a small representative configurations---which incorporates the influences of various configuration options.
Randomized  functions
 select random items~\cite{guo2013variability, sarkar2015cost}. Other, more intricate,
acquisition functions first cluster the data then sample only
a subset of examples within each cluster~\cite{nair17}.
But the more intricate the acquisition function, the longer it
takes to execute---particularly for software with very many configurations.  For example,
recent studies with   a new state-of-the-art acquisition function show that such approaches are limited to models with less than a dozen decisions (i.e. configuration options)~\cite{zuluaga2016varepsilon}.

As an example of an acquisition function, consider
Figure~\ref{fig:bayesian_optimazation}. It illustrates a time series of a typical
 run of SMBO. The bold black line represents the actual performance function ($f$---which is unknown in our setting) and the dotted black line represents the estimated objective function (in the language of SMBO, this is the {\em prior}). 
 The purple regions represent the configuration or uncertainty of estimation in a region---the thicker that region, the higher the uncertainty.
 
 The green line in that figure represents the acquisition function. The optimization starts with two points (t=2). At each iteration, the acquisition function is maximized to determine where to sample next. The acquisition function is a user-defined strategy, which takes into account the estimated performance measures (mean and variance) associated with each configuration.. The chosen sample (or configuration) maximizes the acquisition function (\textit{argmax}). This process terminates when a predefined stopping condition is reached which is related to the budget associated with the optimization process.

Gaussian Process Models (GPM) is often the surrogate model of choice in the machine learning literature.
GPM is a probabilistic regression model which instead of returning a scalar ($f(x)$) returns the mean and variance associated with $x$. There are various acquisition functions  to choose from: (1) Maximum Mean, (2) Maximum Upper Interval, (3) Maximum Probability of Improvement, (4) Maximum Variance, and (5) Maximum Expected Improvement. 

\noindent Building GPMs can be very challenging since:\label{sec:bo_shortcomings}
\bi[leftmargin=*]
\item
GPMs can be very fragile, that is, very sensitive to the parameters of GPMs;
\item
GPMs do not scale to  high dimensional data as well as a large dataset (software system with large configuration space)~\cite{shen2006fast}. For example, in SE, the state-of-the-art in this area using GPMs for  optimization was
limited to models with around ten decisions ~\cite{wang2016bayesian}.
\ei

\section{performance optimization of Configurable Software Systems}
In this section, we discuss the model-based methods used in the prior work to find the better configurations of software systems. \subsection{Residual-based: ``Build an Accurate Model''}\label{sec:residual}
In this section, we discuss the residual-based method for building performance models for software systems, which, in SMBO terminology, is an optimizer with a \textit{flat acquisition function}, that is, all the points are equally likely to be selected (random sampling). 

\begin{figure}[t]
\centering
\includegraphics[scale=0.4]{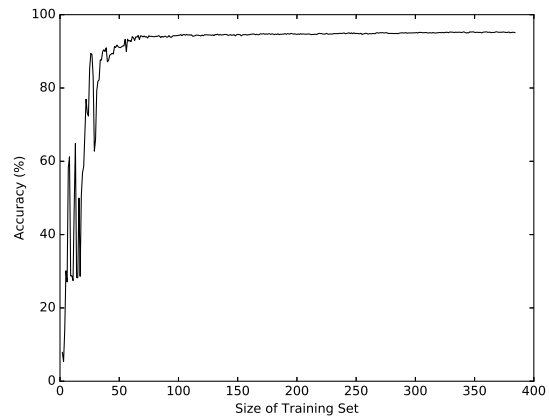}
\caption{{\small The relationship between the accuracy and the number of samples used to train the performance model of the running Word Count application on Apache Storm. Note that the accuracy does not improve substantially after 20 sample configurations.}
}
\label{fig:learning_curve}
\end{figure}

When the cost of collecting data (benchmarking time) is higher than the cost of building a performance model (surrogate model), it is imperative to minimize the number of measurements required for model building. A learning curve shows the relationship between the size of the training set and the accuracy of the model. In Figure~\ref{fig:learning_curve}, the horizontal axis represents the number of samples used to create the performance model, whereas the vertical axis represents the accuracy (measured in terms of MMRE---Mean Magnitude of Relative Error) of the model learned. 
Learning curves typically have a steep sloping portion early in the curve followed by a plateau late in the curve. The plateau occurs when adding data does not improve the accuracy of the model. As engineers, we would like to stop sampling as soon as the learning curve starts to flatten. 
Two types of residual-based methods have been introduced in Sarkar et al. namely \textit{progressive} and \textit{projective sampling}. \\

(1)~\textit{Progressive Sampling} uses an iterative sampling strategy to inform the process of building the performance model. It starts by sampling a small set of configurations and their corresponding performance measures to build a model and validating the model using a holdout set. Configurations are iteratively sampled and used to construct the performance model until the performance model achieves a specified accuracy (measured in terms of MMRE). The sampling process terminates when a predefined threshold is reached.
One of the shortcomings of progressive sampling is that the resulting performance model achieves an acceptable accuracy only after a large number of iterations, which implies high modeling cost. There is no way to determine the cost of modeling until the performance model is already built, which defeats its purpose, as there is a risk of over-shooting the modeling budget and still not obtaining an accurate model. \\
(2)~\textit{Projective sampling} addresses this problem by approximating the learning curve using a minimal set of initial configurations, thus providing the stakeholders with an estimate of the modeling cost.

We use progressive sampling as a representative because projective sampling adds only a sample estimation technique to the progressive sampling and does not add anything to the sampling itself.

The residual-based method discussed here considers only performance configuration optimization scenarios with a single-objective. In the residual-based method, the correctness of the performance model built is measured using error measures such as MMRE:
\begin{equation}
\mathit{MMRE}=\frac{\mid\mathit{f(x)} - \mathit{y}\mid}{\mathit{y}} \cdot 100
\label{eq:err}
\end{equation}
For further details, please refer to Sarkar et. al~\cite{sarkar2015cost}.

Figure~\ref{fig:progressive_sampling} is a generic algorithm that defines the process of progressive sampling. \emph{Progressive} sampling starts by clearly defining the data used in the training set (called training pool) and used for testing the quality of the surrogate model (in terms of residual-based measures) called \textit{holdout set}. The training pool is the set from which the configurations would be selected (randomly, in this case) and then tested against the holdout set. At each iteration, a (set of) data instance(s) of the training pool is added to the training set (Line 9). Once the data instances are selected from the training pool, they are evaluated, which in our setting means measuring the performance of the selected configuration (Line 12). The configurations and the associated performance scores are used to build the performance model (Line 14). The model is validated using the testing set\footnote{The testing data consist of the configurations as well as the corresponding performance scores.}, then the accuracy is computed. In our setting, we assume that the measure is accuracy (higher is better). Once the accuracy score is calculated, it is compared with the accuracy score obtained before adding the new set of configurations to the training set. If the accuracy of the model (with more data) does not improve the accuracy when compared to the previous iteration (lesser data), then life is lost. This termination criterion is widely used in the field of Evolutionary Algorithms to determine the degree of convergence~\cite{krall2015gale}.

\begin{figure}[t]
\small
\hspace{0.4cm}\begin{lstlisting}[xleftmargin=5.0ex,mathescape,frame=none,numbers=left]
  # Progressive Sampling
  def progressive(training, holdout, lives=3): 
    # For stopping criterion
    last_score = -1
    independent_vals = list()
    dependent_vals = list()
    for count in range(1, len(training)):    
      # Add one configuration to the training set
      independent_vals += training[count]      
      # Measure the performance value for the newly
      # added configuration 
      dependent_vals += measure(training_set[count])  
      # Build model
      model = build_model(independent_vals, dependent_vals)      
      # Test Model
      perf_score = test_model(model, holdout, measure(holdout))
      # If current accuracy score is not better than
      # the previous accuracy score, then loose life
      if perf_score <= last_score:
        lives -= 1
      last_score = perf_score
      # If all lives are lost, exit loop
      if lives == 0: break 
    return model
\end{lstlisting}
\caption{\small{Python code of progressive sampling, a residual-based method.}}
\label{fig:progressive_sampling}  
\end{figure}

\subsection{Rank-based: ``Build a Rank-preserving Model''}
As an alternative to the residual-based method, a rank-based method has recently been proposed~\cite{nair2017using}. 
The rank-based method is similar to residual-based method in that it has a \textit{flat acquisition function}, which resembles random sampling. Like the residual-based method, the rank-based method discussed here also considers only performance configuration optimizations with a single-objective. For further details, please refer to Nair et. al~\cite{nair2017using}.

\begin{figure}[t]
\small
\hspace{0.4cm}\begin{lstlisting}[xleftmargin=5.0ex,mathescape,frame=none,numbers=left]
# rank-based method
def rank_based(training, holdout, lives=3): 
  last_score = -1
  independent_vals = list()
  dependent_vals = list()
  for count in range(1, len(training)):  
      # Add one configuration to the training set
      independent_vals += training[count]      
      # Measure the performance value for the newly
      # added configuration 
      dependent_vals += measure(training_set[count])
      # Build model
      model = build_model(independent_vals, dependent_vals)     
      # Predicted performance values
      predicted_performance = model(holdout) 
      # Compare the ranks of the actual performance 
      # scores to ranks of predicted performance scores
      actual_ranks = ranks(measure(holdout))
      predicted_ranks = ranks(predicted_performance)
      mean_RD = RD(actual_ranks, predicted_ranks)
      # If current rank difference is not better than
      # the previous rank difference, then loose life
      if mean_rank_difference <= last_rank_difference:
        lives -= 1
      last_rank_difference = mean_RD
      # If all lives are lost, exit loop
      if lives == 0: break
    return model
      

\end{lstlisting}
\caption{\small{Python code of rank-based method.}
}
\label{fig:rank-based}  
\end{figure}

In a nutshell, instead of using residual measures of errors, as described in Equation~\ref{eq:err}, which depend on residuals ($r = y - f(x)$)\footnote{Refer to Section~\ref{sec:problem_formal} for definitions.}, it uses a rank-based measure. While training the performance model ($f(x)$), the configuration space is iteratively sampled (from the training pool) to train the performance model. Once the model is trained, the accuracy of the model is measured  by sorting the values of $y=f(x)$ from `small' to `large', that is:
\begin{equation}
    f(x_1) \le f(x_2) \le f(x_3) \le ... \le f(x_n).
\end{equation}
The predicted rank order is then compared to the actual rank order. The accuracy is calculated using the mean rank difference ($\mu RD$):
\begin{equation} \label{eq:rank_performance}
    \mathit{\mu RD} = \frac{1}{n} \cdot \mathlarger{\mathlarger{\sum}}_{i=1}^{n}\abs[\Big]{rank(y_i) - rank(f(x_i))}
\end{equation}
This measure simply counts how many of the pairs in the test data have been ordered incorrectly by the performance model $f(x)$ and measures the average of magnitude of the ranking difference.

In Figure~\ref{fig:rank-based}, we list a generic algorithm for the rank-based method. Sampling starts by selecting samples randomly from the training pool and by adding them to the training set (Line 8). Then, the collected sample configurations are evaluated (Line 11). The configurations and the associated performance measure are used to build a performance model (Line 13). The generated model (CART, in our case) is used to predict the performance measure of the configurations in the testing pool (Line 16). Since the performance value of the holdout set is already measured, hence known, the ranks of the actual performance measures, and predicted performance measure are calculated. (Lines 18--19). The actual and predicted performance measure is then used to calculate the rank difference using Equation~\ref{eq:rank_performance}. If the rank difference ($\mu RD$) of the model (with more data) does not decrease when compared to the previous generation (lesser data), then a life is lost (Lines 23--24). When all lives are expired, sampling terminates (Line 27).
\begin{figure}[!t]
\hspace{0.4cm}\begin{lstlisting}[xrightmargin=5.0ex,mathescape,frame=none,numbers=right]
  # ePAL Multi-objective SMBO
  def ePAL(all_configs,  $\varepsilon$, size = 20):  
    # Add |\textit{size}| number of randomly selected 
    # configurations to training data
    train = random.sample(all_configs, size)
    # Measure the performance value for sampled training data
    train = [ measure(x) for x in train ]
    # Remove the measured configurations from configuration space
    all_configs = all_configs $\setminus$ train  
    # Till all the configurations in all_configs has been either sampled or discarded 
    while len(all_configs) > 0:
      # Build GPM
      model = GPM(train)
      # Get prediction and corresponding confidence intervals
      # associated with each configuration in all_configs
      $\mu$, $\sigma$ = model.predict(all_configs)
      # Only keep configurations discard based on uncertainty 
      all_configs = all_configs - discard(all_configs,$\mu$, $\sigma$,  $\varepsilon$)
      # Find and measure another point based on acquisition function.
      new_point = measure(acquisition_function(all_configs, $\mu$, $\sigma$))
      # Add new_point to train 
      train += new_point
    return train 
\end{lstlisting}
\caption{\small{Python code of ePAL, a multi-objective SMBO.}}
\label{fig:ePAL}  
\end{figure}

\subsection{ePAL: ``Traditional SMBO''}\label{sec:epal}

Unlike the residual-based and rank-based methods, epsilon Pareto Active Learning (ePAL) reflects on the evaluated configurations (and corresponding performance measures) to decide the next best configuration to measure using \textit{Maximum Variance} (predictive uncertainty) as an acquisition function.   ePAL incrementally updates a  model (GPM) representing a generalization
of all samples (or configurations) seen so far. 
The model can be used to decide the next most promising configuration to evaluate.  This ability to avoid unnecessary measurement (by just exploring a model)
is very useful in the cases where each measurement can take days to weeks. 

In Figure~\ref{fig:ePAL}, we list a generic algorithm for ePAL. ePAL starts by selecting samples randomly from the configuration space (\textit{all\_configs}) and by adding them to the training set (Line 5). The collected sample configurations are then evaluated (Line 7). The configurations and the associated performance values are used to build a performance model (Line 13). The generated model (GPM, in this case) is used to predict the performance values of the configurations in the testing pool (Line 16). Note that the model returns both the value ($\mu$) as well as the associated confidence interval ($\sigma$). These predicted values are used to discard configurations, which have a high probability of being dominated by another point (Line 18). Domination is defined in Equation~\ref{eq:bdom}~\footnote{ePAL then removes all  $\varepsilon$-dominated points:
$a$ is discarded due to $b$ if $\mu_b + \sigma_b$  $\varepsilon$-dominates $\mu_a - \sigma_a$, where  
$x$  $\varepsilon$-dominates $y$ if $x + \varepsilon \succeq  y$
and  ``$\succeq$'' is   binary domination- see Equation~\ref{eq:bdom}}. After configurations (which have a high probability of being dominated) have been discarded, a new configuration (\textit{new\_point}) is selected and measured (Line 20). The selected configuration \textit{new\_point} is the most uncertain in \textit{all\_configs}. Then, \textit{new\_point} is added to \textit{train}, which is then used to build the model in the subsequent iteration (Line 22). When all the configuration in \textit{all\_configs} have been discarded (or evaluated and moved to train) the process terminates.

Note again, since ePAL is a traditional SMBO, it shares its shortcomings (refer to Section~\ref{sec:bo_shortcomings}).


\section{{\bfseries FLASH}: A Fast Sequential Model-based Method}
 

To overcome the shortcomings of the traditional SMBO, \flash makes certain design choices:
\begin{itemize}[leftmargin=*]
    \item {\sc Flash}\textquotesingle s acquisition function uses  \textit{Maximum Mean}. Maximum Mean returns the sample (configuration) with highest expected (performance) measure;
    \item GPM is replaced with CART~\cite{breiman1984classification}, a fixed-point regression model. This is possible because the acquisition function requires only a single point value rather than the mean and the associated variance. 
\end{itemize}
When used in a multi-objective optimization setting, \flash models each objective as a separate performance (CART) model. This is because the CART model can be trained for one performance measure or dependent value. 

The basic idea of CART is as follows: CART recursively partitions the set of configurations (based on a configuration option) into smaller clusters until the performance of the configurations in the clusters are similar. 
Each split of the set of configurations is driven by a decision on the configuration option that would minimize the entropy or prediction error.
These recursive clustering is represented as a binary decision tree. So, when we need to predict the performance of a new configuration not measured so far, we use the decision tree to find the cluster which is most similar to the new configuration.

\flash replaces the {\em actual} evaluation of all configurations (which can be a very slow process) with a {\em surrogate evaluation}, where the CART decision trees are used to guess
the objective scores (which is a very fast process). Once guesses are made,
then some {\em select} operator must be applied to remove less-than-satisfactory configurations.
Inspired by the
 decomposition approach of MOEA/D~\cite{zhang2007moea}, \flash uses
 the   following 
 stochastic Maximum Mean method, which we call {\em Bazza}\footnote{Short for ``bazzinga''.
Also, ``Bazza'' is Australian for ``Barry''
which the name of Barry Allen of the Flash T.V. series;
 and the childhood nickname of the 44th  United States President Barack Obama.}.

 For problems with $o$ objectives that we seek to maximize,
 {\em Bazza} returns the configuration that has outstandingly maximum objective values across $N$ random projections. Using the {\em predictions}
for the $o$ objectives from the learned CART models,
then {\em Bazza} executes as follows.  Note that the first step (randomly assigning weights to goals)
is a technique we burrow and adapt from the MOEA/D algorithm~\cite{zhang2007moea}.

\bi[leftmargin=*]
 \item
$N$ vectors $V$ of length $m$ are generated and filled with random numbers of the range 0..1. This represents the various weight vectors.  The idea is to decompose one problem into a set of $N$ sub-problems (uniformly spread $N$ weight vectors.
\item Set $\mathit{max=0}$ and $\mathit{best=nil}$.
  \item For each configuration $x_i$
 \bi
 \item
 Guess its performance scores $y_{i,j}$ using the {\em predictions} from the CART models.
\item
Compute its mean weight as follows: 
\begin{equation}\label{eq:mm}
 \centering
 mean_i=\frac{1}{N}\sum_n^N \sum_j^{{\mathit m}}\left(V_{n,j} \cdot x_{i,j}\right)
 \end{equation}
 \item
 If $\mathit{mean} > \mathit{max}$, then  $\mathit{max:=mean}$ and $\mathit{best}:=x_i$.
 \ei
\item
Return $\mathit{best}$.
\ei

In summary, given a set of $V$ weight vectors of length $m$,  {\em Bazza}  finds
the vector that scores best across  $N$ different weighted sums, each of which is computed with random weight vectors.

\begin{figure}
\small
\hspace{0.2cm}\begin{lstlisting}[xrightmargin=5.0ex,mathescape,frame=none,numbers=right]
def FLASH(uneval_configs, fitness, size, budget):
  # Add |size| number of randomly selected configurations to training data. 
  # All the randomly selected configurations are measured
  eval_configs = [measure(x) for x in sample(uneval_configs, size)]
  # Remove the evaluations configuration from data
  uneval_configs.remove(eval_configs)
  # Till all the lives has been lost
  while budget > 0:
    # build one CART model per objective
    for o in objectives: model[o] = CART(eval_configs)
    # Find and measure another point based on acquisition function
    acquired_point = measure(acquisition_fn(uneval_configs, model))
    eval_configs += acquired_point  # Add acquired point 
    uneval_config -= acquired_point # Remove acquired point 
    # Stopping Criteria
    budget -= 1
  return best
  
def acquisition_fn(uneval_configs, model, no_directions=10):  
  # Predict the value of all the unevaluated configurations using model
  predicted = model.predict(uneval_configs)
  # If number of objectives is greater than 1 (In our setting len(objectives) = 2)
  if len(objectives) > 1: # For multi-objective problems
    return  Bazza( predicted )
  else:  # For single-objective problems
    return max(predicted)  
\end{lstlisting}
\caption{Python code of \flash}
\label{fig:flash_frame}  
\end{figure}

The resulting algorithm is shown in  Figure~\ref{fig:flash_frame}. 
Before initializing \flash, a user needs to define three parameters \textit{N}, \textit{size}, and \textit{budget} (refer to Section~\ref{sec:parameter_tuning} for a sensitivity analysis). \flash starts by randomly sampling a predefined number (\textit{size}) of configurations from the configuration space and evaluate them (Line 4). The evaluated configurations are removed from the unevaluated pool of configurations (Line 6). The evaluated configurations and the corresponding performance measure/s are then used to build CART model/s (Line 10). This model (or models, in case of multi-objective problems) is then used by the acquisition function to determine the next point to measure (Line 13). The acquisition function accepts the model (or models) generated in Line 10 and the pool of unevaluated configurations (\textit{uneval\_configs}) to choose the next configuration to measure. The model is used to generate a prediction for the unevaluated configurations. 
For single-objective optimization problems, the next configuration to measure is the configuration with the highest predicted performance measure (Line 26). For multi-objective optimization problems, 
{\em Bazza} is applied. The configuration chosen by the acquisition function is evaluated and added to the evaluated pool of configurations (Line 12-13) and removed from the unevaluated pool (Line 14). \flash terminates once it runs out of budget (Line 8).





Note the advantages of this approach:
SMBO is a widely used method~\cite{hoffman2014modular}  for many important tasks (cloud configuration~\cite{alipourfard2017cherrypick}, hyperparameter optimization~\cite{bergstra2011algorithms}) so even a small improvement in this method would be significant for a large of number of domains
Truly \flash includes many novel innovations. The following list shows the significant innovations of this work and the delta to prior research.  \\
\textbf{1.} Evolutionary algorithms (EAs) have been used for optimizing black-box optimization~\cite{harman2009search}. Using Evolutionary Algorithms is relatively straightforward since it requires no domain knowledge to solve a problem.\\
\noindent\textbf{Challenge:} Evolutionary algorithms suffer from two problems.
Firstly, there is the issue of
the number of evaluations required for an EA. A standard EA experiment is 100 individuals mutated for 100+ generations~\cite{sarro2016multi}. This renders Evolutionary Algorithms unsuitable for  our domain since individual evaluation can be very slow (requires re-running a benchmark suite).
A second problem with EAs is the problem of slow convergence (i.e., the performance delta across these generations may be very slow and take a long time to stabilize ~\cite{chen2017beyond}).
For this reason, research in this area in the last decade has explored non-EA methods for software configuration~\cite{siegmund2012predicting, guo2013variability, sarkar2015cost, nair17, nair2017using}.\\
\noindent\textbf{New approach:} 
\bi
\item
Here we explore SMBO for software configuration optimization. 
\item
 While SMBO is gaining some popularity in other domains~\cite{alipourfard2017cherrypick, golovin2017google, shahriari2016taking },  this article is the first reporting a successful application of SMBO to software configuration.
\ei
\textbf{2.} Prior work in this area used some surrogate model learned by data mining (e.g., with CART ~\cite{guo2013variability, sarkar2015cost, nair17}),   possibly combined with random sampling. Such surrogates are useful for guiding the construction of better configuration models since they can be much faster to execute than (say) re-running a benchmark. Hence, an optimizer that uses such surrogates can terminate relatively quickly.\\
\noindent\textbf{Challenge:} One drawback with  surrogate models is that they require a holdout set, against which the surrogate model (built iteratively) is evaluated. Interestingly, prior work does not discuss the cost of populating, which may require exploring up to  20\% of the total configuration space~\cite{nair2017using}.\\
\noindent\textbf{New approach:} 
\bi
\item
We found that applying SMBO removes the need for this hold out. We build the model incrementally, thus the configurations (and their performance) sampled at a given point in time used to intelligently select the next data point to collect.
\item
This work is the first successful application of such incremental model construction (with no holdout sets)  for software configuration.
\ei
\noindent \textbf{3.} Standard SMBO algorithms are a widely used method for finding good samples (as used in hyper-parameter optimization). Standard SMBO builds its models using a method called  Gaussian Process Models. Due to internal complexities of some of its matrix operations, GPM can  handle only a dozen configuration options (or less), while modern software may required many more configuration options. \\
\noindent\textbf{Challenge:} How can we scale SMBO to much larger configuration options?\\
\noindent\textbf{New approach:} 
\bi
\item
One of the core innovations of this article is the use of CART (one CART per goal) for surrogate modeling. That is, we replace GPM with CART.  GPM takes time O($M^3$)~\cite{rasmussen2004gaussian} while recursive bifurcating algorithms like CART are much faster (takes time O($MN^2$) to build its trees where M is the size of the training dataset and N is the number of attributes~\cite{su2006fast}). Furthermore, methods such as  CART have been extensively studied, and very fast incremental versions are simple to implement---see Domingos et al.~\cite{domingos2000mining}, which is to say that if CART ever gets slow, there are many alternatives, we could try that would readily speed it up. 
\item
Theoretical complexity results aside, we demonstrate empirically that that our  CART-based method scales much better than GPM. As of 2016,  published state-of-the-art results~\cite{wang2016bayesian} report that they were unable to use more than 10 attributes within a GPM. As of 2018, our own experiments confirm that GPM cannot scale to more than a dozen configuration options. As shown in Figures~\ref{fig:multiconfig_time}(a) and (b),  \flash  scales linearly in number of attributes to models that defeats SMBO.
We regard FLASH\textquotesingle s ability to scale linearly in number of attributes to be a major contribution  of this article.
\ei
\textbf{4.} One   advantages of SMBO algorithms such as GPM is that they identify   region(s) of most variance within a model. Such regions   represent zones of most uncertainty (and sampling there has greatest chance of most improving a model). \\
\noindent\textbf{Challenge:} If we are not using GPM, how can we find the best regions for future sampling?\\
\noindent\textbf{New approach:}
\bi
\item
Another  core innovation is the \textit{Bazza} algorithm.   \textit{Bazza}   assumes that the greatest mean might contain the values that most extend to the desired maximal (or minimal) goals.   \textit{Bazza} finds that region in
 linear time (since it only has to track the most extreme values seen so far). 
\item
\textit{Bazza} is an important innovation since standard methods for finding the best candidates within a population of size M require time  O($M^2$)~\cite{deb2002fast}. But as shown in this article, our methods require only O(M).
\ei
\noindent5. The kernels used in Gaussian Process Model assume ``smoothness''~\cite{rasmussen2004gaussian}, or in other words, the configurations which are closer to each other have similar performance. In the case of software configuration, this assumption is highly unlikely since we know of many software options where a small change can lead to  radically different software performance
(e.g. switching from link lists to  B-trees is a single change to  one value of one configuration option---but that change can lead to dramatic speed ups in the software).
\noindent\textbf{Challenge:} How to avoid  GPM's smoothness assumptions?\\
\noindent\textbf{New approach:} 
\bi
\item
We use CART, a learner that recursively  bifurcates training data into different regions. The important point  here is that CART makes no assumption that neighboring regions have the same properties.
\item
Unlike prior work, our use of CART makes no limiting  assumptions about the smoothness of the space.
\ei

\section{Evaluation}

\subsection{Research Questions}

In prior work, performance configuration optimization was conducted by sequentially sampling configurations to build models, both accurate (residual-based method) and inaccurate (rank-based method). 
Both methods divide the configuration space into (i) training pool, (ii) validation set, and (iii) holdout set. They sequentially select a configuration from the training pool and add it to the training set (which is a subset of the training pool). The configuration (along with the corresponding performance measure) is used to build a model. Both methods use a validation set to evaluate the quality of the model. The size of the validation set is based on an engineering judgment and expected to be a representative of the whole configuration space. Prior work~\cite{nair2017using} used 20\% of the configuration space as holdout set, but did not consider the cost of using the validation set.

Our research questions are geared towards assessing the performance of \flash based on two aspects: (i) \textbf{Effectiveness} of the solution or the rank-difference between the best configuration found by \flash to the actual best configuration, and (ii) \textbf{Effort} (number of measurements) required to find the better configuration.

\noindent The above considerations lead to two research questions:
\noindent\textbf{RQ1}\textit{: Can  \flash find the better configuration?}\\
Here, the better configurations found using \flash are compared to the ones identified in prior work, using the residual-based and rank-based method. The effectiveness of the methods is compared using rank-difference (Equation~\ref{eq:rankdiff}).

\noindent\textbf{RQ2}\textit{: How expensive is \flash (in terms of
 how many configurations must be measured)?}\\
It is expensive to build (residual-based or rank-based) models since they require using a holdout set. Our goal is to  demonstrate that \flash can find better configurations of a software system using fewer measurements.

To the best of knowledge, SMBO has never been used for multi-objective performance configuration optimization in software engineering. However, similar work has been done by Zuluaga et al.~\cite{zuluaga2016varepsilon} in the machine learning community, where they introduced ePAL. We use ePAL as a state-of-the-art method to compare \flash.

We do not consider the work by Oh et al.~\cite{oh2017finding}, which uses true random sampling to find the better configurations. We do not compare \flash with Oh et al.\textquotesingle s method mainly for the following reason: Oh et al.\textquotesingle s work supports only Boolean configuration options, which limits its practical applicability. \flash, and the prior work considered in this article, do not have this limitation. Moreover, Oh\textquotesingle s work is limited to single-objective problems. 
Since it does not build a performance model during the search process, it cannot be easily adapted  to multi-objective problems. 
One may argue that running Oh et al.\textquotesingle s approach alternatively on different objectives (of a multi-objective problem) could lead to a set of solutions on the Pareto Front. However, this is not a proper alternative since these runs (on separate objectives) are independent of each other (i.e., they are run separately and cannot inform each other).


Since ePAL suffers from the shortcomings of traditional SMBO, our research questions are geared towards finding the estimated Pareto-optimal solutions (predicted Pareto Frontier~\footnote{Pareto Frontier is a set of solutions which are non-dominated by any other solution.}), which is \textit{closest} to the true Pareto Frontier (which requires measuring all configurations) with \textit{least effort}.  We assess the performance of \flash by considering three aspects: (i) \textbf{Effectiveness} of the configurations between the Pareto Frontier and the ones approximated by an optimizer, and \textbf{Effort} evaluated in terms of (ii) number of measurements, and (iii) time to approximate the Pareto Frontier.

\noindent The above considerations lead to three research questions:
\textbf{RQ3:} \textit{How effective is \flash for multi-objective performance configuration optimization?}\\
The effectiveness of the solution or the difference between the predicted Pareto Frontier found by optimizers to the true Pareto Frontier,\\
\textbf{RQ4:} \textit{Can \flash be used to reduce the effort of multi-objective performance configuration optimization compared to ePAL?}\\
Effort  (number of  measurements)  required to estimate the Pareto Frontier which is closest to the true Pareto Frontier, and\\
\textbf{RQ5} \textit{Does \flash save time for multi-objective performance configuration optimization compared to ePAL? }\\
Since ePAL may take substantial time to find the approximate the Pareto Frontier, it is imperative to show that \flash can approximate the Pareto Frontier and converge faster.

Our goal is to minimize the effort (time and number of measurements) required to find an approximate Pareto Frontier as close to the actual Pareto Frontier as possible.

\subsection{Case Studies}

We evaluated \flash in two different types of problems namely: (1) single-objective optimization problems and (2)  multi-objective optimization problems using 30 scenarios (15 scenarios in multi-objective settings) from 6 software systems. These systems are summarized in Table~\ref{fig:software_systems}.
More details about the software systems are available at \url{http://tiny.cc/flash_systems/}.

We selected these software systems since they are widely used in the configuration and search-based SE
literature~\cite{siegmund2012predicting, guo2013variability, sarkar2015cost, nair17, nair2017using, oh2017finding, jamshidi2016uncertainty, zuluaga2016varepsilon} as benchmark problems for this kind of optimization work. Furthermore, extensive documentation
is available at the supplementary Web site for all these models.

\subsection{Experimental Rig}
 
 \subsubsection{Exploring RQ1, RQ2}
 
For each subject system, we build a table of data, one row per valid configuration. We then run all configurations of all systems (that is, that are invoked by a benchmark) and recorded the performance scores. 
Note that, while answering the research questions, we ensure that we never test any prediction model on the data that we used to learn the model. 

To answer our research questions, we split the datasets into training pool (40\%), holdout set (20\%), and validation pool (40\%). The size of the holdout set is taken from prior work~\cite{nair2017using}. It is worth to note that this is a hyper-parameter and is set based on an engineering judgment. To perform a fair comparison while comparing \flash with prior work, the training pool and validation pool are merged for \flash experiments.

The experiment to find better configuration using the residual-based and rank-based methods is conducted in the following way:
\begin{itemize}[leftmargin=*]
    \item Randomize the order of rows  in  the training data
    \item \textbf{Do}
    \begin{itemize}
        \item Select one configuration (by sampling with replacement) and add it to the training set
        \item Determine the performance scores associated with the configuration. This corresponds to a table look up but would entail compiling or configuring and executing a system configuration in a practical setting.
        \item Using the training set and the accuracy, build a performance model using CART.
        \item Using the data from the testing pool, assess the accuracy either using MMRE (as described in Equation~\ref{eq:err}) or rank difference (as described in Equation~\ref{eq:rank_performance}).         
    \end{itemize}
    \item \textbf{While} the accuracy is greater or equal to the threshold determined by the practitioner (rank difference in the case of rank-based method and MMRE in the case of residual-based method).
\end{itemize}
Once the model is trained, it is tested on the data in the validation pool. Please note, the learner has not been trained on the validation pool.
The experiment to find better configuration by \flash is conducted in the following way:
\begin{itemize}[leftmargin=*]
    \item Choose 80\% of the data (at random)\footnote{We use 80\% because other method find the better configuration sampling from a training set of 40\% and test is against a testing pool of 40\%. To make sure, we make a fair comparison, we use \flash to find the best configuration among 80\% of the configuration space.}
    \item Randomize the order of rows  in  the training data
    \item \textbf{Do}
    \begin{itemize}
    \item Select 30 configurations (by sampling with replacement) and add them to the training set
        \item Determine the performance scores associated with the configurations. This corresponds to  a table look up, but would entail compiling or configuring and executing a system configuration in a practical setting.
        \item Using the training set, build a performance model using CART.
        \item Using the CART model, find the configuration with best predicted performance.
        \item Add the configuration with best predicted performance to the training set.
    \end{itemize}
    \item \textbf{While} the stopping criterion (\textit{budget}) is not met, continue adding configurations to the training set.
\end{itemize}

Once \flash has terminated, the configuration with the best performance is selected as the better configuration. Please note that unlike the methods proposed in prior work, there is no training and validation pool in \flash. It uses the whole space and returns the configuration with the best performance.

\textbf{RQ1} relates the results found by \flash to ones of residual-based and rank-based methods. We use the absolute difference between the ranks of the configurations predicted to be the best configuration and the actual optimal configuration. We call this measure rank difference.
\begin{equation}\label{eq:rankdiff}
    \begin{split}
        RD &= \abs[Big]{\mathrm{rank(actual_{best})} - \mathrm{rank(predicted_{best})}}\\
    \end{split}
\end{equation}

\noindent Ranks are calculated by sorting the configurations based on their performance scores. The configuration with the minimum performance score, $\mathrm{rank(actual_{best})}$, is ranked 1 and the one with the highest score is ranked as $N$, where $N$ is the number of configurations.

\begin{figure*}[t]
\centering
\includegraphics[width=0.85\paperwidth, height=4.7cm]{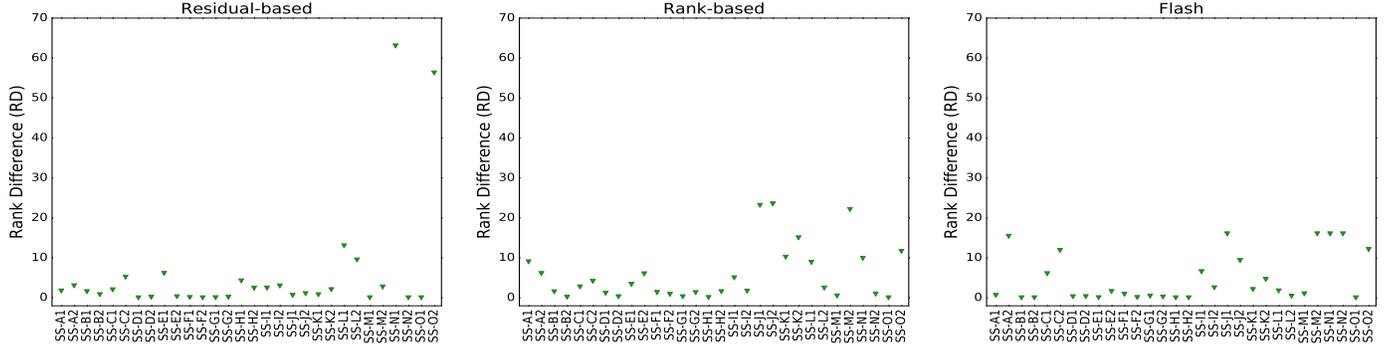}
\caption{
    \small{The rank difference of the prediction made by model built using the residual-based method, the rank-based methods, and \flash. Note that the y-axis of this chart rises to large values; e.g., SS-M has 239,260 possible configurations. Hence, the above charts could be summarized as follows: ``the \flash is surprisingly accurate since the rank difference is usually close to 0\% of the total number of possible configurations.'' 
    }
}\label{fig:one_obj_rd}
\end{figure*}

\subsubsection{Exploring RQ3, RQ4, and RQ5}\label{sec:exp_mo}
Similar to \textbf{RQ1} and \textbf{RQ2}, for each subject system, we build a table of data, one row per valid configuration. We then run all configurations of all systems and record the performance scores. To this table, we add two columns of measurements (one for each objective) obtained from measurements.

To measure effectiveness, we use quality indicators as advised by Wang et al.~\cite{wang2016practical}. The quality indicators are:
\bi[leftmargin=*]
\item
The \textit{Generational Distance} (GD)~\cite{van1999multiobjective} measures the closeness  of the solutions from by the optimizers to the 
{\em Pareto frontier} that is, the actual set of non-dominated solutions. 
\item
The \textit{Inverted Generational Distance} (IGD)~\cite{coello2004study}
is the mean distance from points on the  \textit{true} Pareto-optimal solutions to its nearest point in the predicted Pareto-optimal solutions returned by the optimizer.
\ei
Note that, for both measures, {\em smaller} values are {\em better}.
Also,
according to
Coello et al.~\cite{coello2004study}, IGD is a better measure of how well solutions of a method are {\em spread}   across the space of all known solutions. A lower value of GD indicates that the predicted Pareto-optimal solutions have converged (or are near) to the actual Pareto-optimal solutions. However, it does not comment on the diversity (or spread) of the solutions. GD is useful while interpreting the results of \textbf{RQ3} and \textbf{RQ6}, where we would notice that \flash has low GD values but relatively high IGD values.

To answer our research questions, we initialize ePAL and \flash with randomly selected configurations along with their corresponding performance scores. Since, ePAL does not have an explicit stopping criterion, we allow ePAL to run until completion. For \flash, we allowed a budget of 50 configurations. The value 50 was assigned by parameter tuning (from Section~\ref{sec:parameter_tuning}). The configurations evaluated during the execution of the three methods are then used to measure the quality measures (to compare methods).  Note that we use two versions of ePAL: ePAL with $\epsilon=0.01$ (ePAL\_0.01), and ePAL with $\epsilon=0.3$ (ePAL\_0.3)~\footnote{Refer to Section~\ref{sec:epal} for definition of $\epsilon$}. These ePAL versions represents two extremes of ePAL from the most cautious ($\epsilon=0.01$)---maximizing quality to most careless ($\epsilon=0.3$)---minimizing measurements.~\footnote{We have measured other values of epsilon between 0.01 and 0.3, but due to space constraints we show results from two variants of ePAL}

Other aspects of our experimental setting were designed in response to the specific features of the experiments. For example, all the residual-based, rank-based and \flash methods are implemented in Python. We use Zuluaga et al.\textquotesingle s implementation of ePAL, which was implemented in Matlab. Since we are comparing methods implemented in different languages, we measure ``speed'' in terms of the number of measurements (a language-independent feature) along with runtimes.

\begin{figure}[tbh]
    {
{\scriptsize \begin{tabular}{l@{~~~}l@{~~~}r@{~~~}r@{~~~}c}
\rowcolor{lightgray}\arrayrulecolor{lightgray}
\textbf{SS-A2} & \textbf{} & \textbf{} & \textbf{} & \\\hline
  1 &   Rank-based &    1.0  &  4.0 & \quart{0}{15}{3} \\
  1 & Residual-based &    2.0  &  5.0 & \quart{0}{19}{7} \\
\hline  2 &        Flash &    9.0  &  18.0 & \quart{7}{72}{35} \\
\hline \end{tabular}}

{\scriptsize \begin{tabular}{l@{~~~}l@{~~~}r@{~~~}r@{~~~}c}
\rowcolor{lightgray}\arrayrulecolor{lightgray}
\textbf{SS-C1} & \textbf{} & \textbf{} & \textbf{} & \\\hline
  1 & Residual-based &    0.0  &  2.0 & \quart{0}{22}{0} \\
  1 &   Rank-based &    1.0  &  3.0 & \quart{0}{34}{11} \\
\hline  2 &        Flash &    3.0  &  6.0 & \quart{11}{68}{34} \\
\hline \end{tabular}}

{\scriptsize \begin{tabular}{l@{~~~}l@{~~~}r@{~~~}r@{~~~}c}
\rowcolor{lightgray}\arrayrulecolor{lightgray}
\textbf{SS-C2} & \textbf{} & \textbf{} & \textbf{} & \\\hline
  1 &   Rank-based &    2.0  &  2.0 & \quart{0}{9}{4} \\
  1 & Residual-based &    2.0  &  5.0 & \quart{0}{24}{4} \\
\hline  2 &        Flash &    7.0  &  16.0 & \quart{0}{79}{29} \\
\hline \end{tabular}}

{\scriptsize \begin{tabular}{l@{~~~}l@{~~~}r@{~~~}r@{~~~}c}
\rowcolor{lightgray}\arrayrulecolor{lightgray}
\textbf{SS-E2} & \textbf{} & \textbf{} & \textbf{} & \\\hline
  1 & Residual-based &    0.0  &  0.0 & \quart{0}{0}{0} \\
\hline  2 &   Rank-based &    1.0  &  2.0 & \quart{0}{79}{39} \\
  2 &        Flash &    1.0  &  1.0 & \quart{0}{39}{39} \\
\hline \end{tabular}}

{\scriptsize \begin{tabular}{l@{~~~}l@{~~~}r@{~~~}r@{~~~}c}
\rowcolor{lightgray}\arrayrulecolor{lightgray}
\textbf{SS-J1} & \textbf{} & \textbf{} & \textbf{} & \\\hline
  1 & Residual-based &    0.0  &  1.0 & \quart{0}{1}{0} \\
\hline  2 &        Flash &    1.0  &  13.0 & \quart{0}{24}{1} \\
  2 &   Rank-based &    13.0  &  42.0 & \quart{1}{78}{24} \\
\hline \end{tabular}}

{\scriptsize \begin{tabular}{l@{~~~}l@{~~~}r@{~~~}r@{~~~}c}
\rowcolor{lightgray}\arrayrulecolor{lightgray}
\textbf{SS-J2} & \textbf{} & \textbf{} & \textbf{} & \\\hline
  1 & Residual-based &    0.0  &  1.0 & \quart{0}{4}{0} \\
\hline  2 &   Rank-based &    2.0  &  14.0 & \quart{0}{69}{9} \\
  2 &        Flash &    4.0  &  16.0 & \quart{0}{79}{19} \\
\hline \end{tabular}}

{\scriptsize \begin{tabular}{l@{~~~}l@{~~~}r@{~~~}r@{~~~}c}
\rowcolor{lightgray}\arrayrulecolor{lightgray}
\textbf{SS-K1} & \textbf{} & \textbf{} & \textbf{} & \\\hline
  1 & Residual-based &    0.0  &  1.0 & \quart{0}{26}{0} \\
\hline  2 &   Rank-based &    1.0  &  3.0 & \quart{0}{79}{26} \\
  2 &        Flash &    1.0  &  3.0 & \quart{0}{79}{26} \\
\hline \end{tabular}}

}
    
    \caption{
        {\small 
            The median rank difference of 20 repeats. Median ranks are the rank difference as described in Equation~\ref{eq:rankdiff}, and IQR the difference between 75th percentile and 25th percentile found during multiple repeats. 
            Lines with a dot in the middle 
            (~\protect\quartex{-2}{13}{6}), 
            show the median as a round dot within the IQR.
            All the results are sorted by the median rank difference: a lower median value is better. 
            The left-hand column (\textit{Rank}) ranks the various methods for example when comparing three techniques. For SS-A2, a rank-based method has a different rank since their median rank difference is statistically different. Please note, this chart only shows software systems where \flash is significantly worse than methods from prior work.
        }
    }
    \label{fig:stat-test}
\end{figure}

\begin{figure*}[t]
        \centering
        \includegraphics[width=0.7\paperwidth, height=6.3cm]{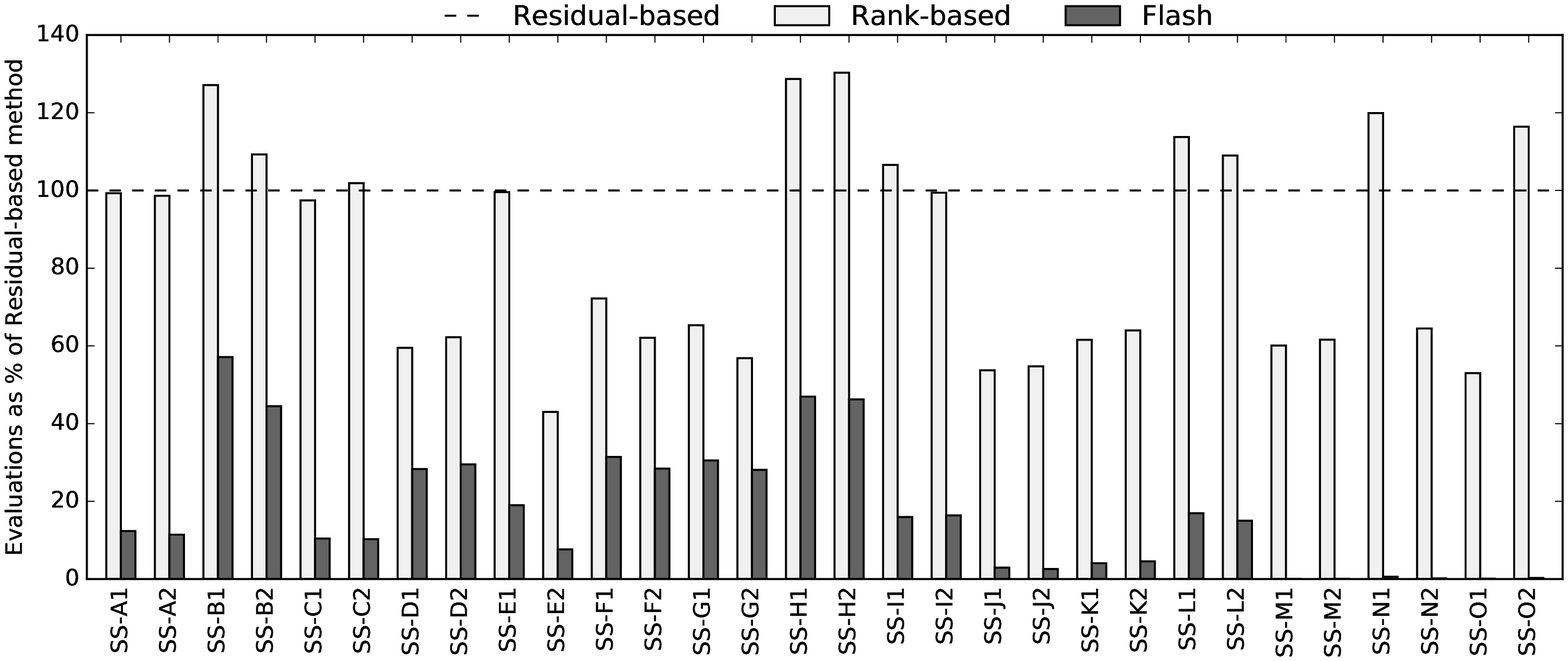}

        \caption{
        \small{
        Measurements required to find better configurations with the residual-based method as the baseline.\vspace{5mm}}
        }\label{fig:evals}
\end{figure*}

\section{Results}
\subsection{Single-objective Problems}\label{sec:so_results}
\noindent\textbf{RQ1: Can  \flash find the better configuration?}\\
Figure~\ref{fig:one_obj_rd} shows the \textit{Rank Difference} of the predictions using \flash, the rank-based method and the residual based method. 
The horizontal axis shows subject systems. The vertical axis shows the rank difference (Equation~\ref{eq:rankdiff}).
\begin{itemize}[leftmargin=*]
    \item 
    The ideal result would be when all the points lie on the line \textit{y=0} or the horizontal axis, which means the method was able to find the better configurations for all subject systems or the rank difference between the predicted optimal solution and the actual optimal solution is 0. 
    \item The sub-figures (left to right) represent the residual-based method, rank-based method, and \flash.
\end{itemize}

\noindent Overall, in Figure~\ref{fig:one_obj_rd}, we find that:
\begin{itemize}[leftmargin=*]
    \item All methods can find better configurations. For example, \flash for SS-J1 predicted the configuration whose performance score is ranked 20th in configuration space. That is \textit{good enough} since \flash finds the 20th most performant configuration among 3072 configurations.
    \item The mean rank difference of the predicted optimal configuration is 6.082, 5.81, and 5.58~\footnote{The median rank difference is 1.61, 2.583, and 1.28.} for residual-based, rank-based, and \flash. 
\end{itemize}
So, the rank of the better configuration found by all the three methods is practically the same. To verify the similarity is statistically significant, we further studied the results using non-parametric tests Scott-Knott test  recommended by Mittas and Angelis~\cite{mittas13} and Arcuri \& Briand~\cite{mittas13}.

Scott-Knott is a top-down clustering approach used to rank different treatments. If the approach finds an interesting division of the data, then some 
statistical test is applied to the two divisions to check if they are 
statistically significantly different. If so, Scott-Knott recurses into both 
halves.
To  apply Scott-Knott, we sorted a list of  $l=20$ values of performance of different method found by different methods. Then, we split $l$ into 
sub-lists $m,n$ to maximize the expected value of differences in 
the observed performances before and after division. For example, for lists 
$l,m,n$ of size $ls,ms,ns$ where $l=m\cup n$: 
\[E(\Delta)=\frac{ms}{ls}|m.\mu - l.\mu|^2 + \frac{ns}{ls}|n.\mu - 
l.\mu|^2\] 
We then apply a   statistical hypothesis test $H$ to check
if $m,n$ are significantly different  (in our case, the conjunction of A12 
and bootstrapping). If so, Scott-Knott recurses on the splits. In other 
words, we divide the data if \textit{both} bootstrap sampling and effect 
size test agree that a division is statistically significant (with a 
confidence level of 99\%) and not a small effect ($A12 \ge 0.6$).

For a justification of the use of non-parametric bootstrapping, see Efron 
\& Tibshirani~\cite[p220-223]{efron93}. For a justification of the use of 
effect size tests, see Shepperd and MacDonell~\cite{shepperd12a}; 
Kampenes~\cite{kampenes07}; and Kocaguenli et 
al.~\cite{Kocaguneli2013:ep}. These researchers warn that, even if a 
hypothesis test declares two populations to be ``significantly'' 
different, then that result is misleading if the ``effect size'' is very 
small. Hence, to assess the performance differences we first must rule out 
small effects using A12.

In Figure~\ref{fig:stat-test}, we show the Scott-Knott ranks for the three methods. The quartile charts show the Scott-Knott results for the subject systems, where \flash did not do as well as the other two methods. For example, the statistic test for SS-C2 shows that the rank difference of configurations found by \flash is statistically larger from the other methods. This is reasonably close, since the median rank of the configurations found by \flash is 7 of 1512 configurations, where for the other methods found configurations have a median rank of 2. As engineers, we feel that this is close because we can find the 7th best configuration using 34 measurements compared to 339 and 346 measurements used by other methods. Overall, our results indicate that:
\vskip 1ex
 \begin{myshadowbox}
         \flash can find better configurations, similar to the residual-based and the rank-based method, of a software system without using a holdout set. 
 \end{myshadowbox}

\noindent\textbf{RQ2:  How expensive is \flash (in terms of how many configurations must be executed)?}\\
To recommend \flash as a cheap method for performance optimization, it is important to demonstrate that it requires fewer measurements to find the better configurations. In our setting, the cost of finding the better configuration is quantified by number of measurements required (i.e., table lookup). Figure~\ref{fig:evals} shows our results. 
The vertical axis represents the ratio of the measurements of different methods
are represented as the percentage of number of measurements required
by residual-based method – since it uses the most measurements
in 66\% scenarios.

Overall, we see that \flash requires the least number of measurements to find better configurations. For example, in SS-E1, \flash requires 9\% of the measurements when compared with the residual-based method and the rank-based method.  There are few cases (SS-M1 to SS-O2) where \flash requires less than 1\% of the residual-based method, which is because these systems have a large configuration space and the holdout set required by the residual-based method and the rank-based method (except \flash) uses 20\% of the measurements. 
\vskip 1ex
 \begin{myshadowbox}
        For performance configuration optimization, \flash is cheaper than the state-of-the-art method. In 57\% of the software systems, \flash requires an order of magnitude fewer measurement compared to the residual-based method and rank-based method.  
 \end{myshadowbox}

\subsection{Multi-objective Optimization}\label{sec:mo_results}

\noindent\textbf{RQ3: How effective is \flash for multi-objective performance configuration optimization?} \\
Table~\ref{table:multi-config} shows the results of a statistical analysis that compares the quality measures of the approximated Pareto-optimal solutions generated by \flash to those generated by ePAL.
\begin{itemize}[leftmargin=*]
    \item The rows of the table shows median numbers of 20 repeated runs over 15 different scenarios.
    \item The columns report the quality measures, generational distance (GD), and inverted generation distance (IGD). Recall \textit{smaller} values of GD and IGD are \textit{better}.
    \item `\textbf{X}' denotes cases where a method did not terminate within a reasonable amount of time (10 hours).
    \item \textbf{Bold}-typed results are statistically better than the rest.
    \item The last row of the table (Win (\%)) reports the percentage of times a method is significantly better than other methods overall software systems.
\end{itemize}

One way to get a quick summary of this table is to read the last row (Win(\%)). This row is the percentage number of times a method was marked statistically better than the other methods. From Table~\ref{table:multi-config}, we can observe that \flash outperforms variants of ePAL since \flash has the highest Win(\%) in both quality measures. This is particularly true for scenarios with more than 10 configuration options, where ePAL failed to terminate while \flash always did. 

We further notice that ePAL-0.01 has a higher win percentage than ePAL-0.3. This is not surprising since (as discussed in Section~\ref{sec:exp_mo}) ePAL-0.01 (optimized for quality) is more cautious than ePAL-0.3 (which is optimized for speed measured in terms of number of measurements). This can be regarded as a sanity check.  It is interesting to note that \flash has impressive convergence score (lower GD scores)---it converges better for 93\% of the systems, but not so remarkable in terms of the spread (lower IGD scores). However, the performance of \flash is similar to ePAL.
It is also interesting that, for software systems where \flash was not statistically better, these are cases where the statistically better method always converged to the actual Pareto Frontier (with few exceptions). 

\vskip 1ex
 \begin{myshadowbox}
         \flash is effective for multi-objective performance configuration optimization. It also works in software systems with more than 10 configuration options whereas ePAL does not terminate in  reasonable time.
 \end{myshadowbox}

\begin{table*}[]
\centering
\caption{Statistical comparisons of \flash and ePAL regarding the Performance measures are GD
(Generational Distance), IGD (Inverted Generational Distance) and a number of measurements. For all measures, less is better; “X” denotes cases where methods did not terminate within a reasonable amount of time (10hrs). The numbers in bold represent statistically better runs than the rest. For example, for SS-G, GD of \flash is statistically better than of ePAL.}\label{table:multi-config}
\begin{tabular}{@{}lrrrrrrrrr@{}}
\toprule
\multirow{2}{*}{\textbf{Software}} & \multicolumn{3}{c}{\textbf{GD}} & \multicolumn{3}{c}{\textbf{IGD}} & \multicolumn{3}{c}{\textbf{Evals}} \\ \cmidrule(l){2-10} 
 & \multicolumn{1}{l}{\textbf{epal\_0.01}} & \multicolumn{1}{l}{\textbf{epal\_0.3}} & \multicolumn{1}{l}{\textbf{\flash}} & \multicolumn{1}{l}{\textbf{epal\_0.01}} & \multicolumn{1}{l}{\textbf{epal\_0.3}} & \multicolumn{1}{l}{\textbf{\flash}} & \multicolumn{1}{l}{\textbf{epal\_0.01}} & \multicolumn{1}{l}{\textbf{epal\_0.3}} & \multicolumn{1}{l}{\textbf{\flash}} \\ \midrule
\textbf{SS-A} & \textbf{0.002} & \textbf{0.002} & \textbf{0} & \textbf{0.002} & 0.002 & \textbf{0} & 109.5 & 73.5 & \textbf{50} \\
\textbf{SS-B} & \textbf{0} & \textbf{0} & \textbf{0.005} & \textbf{0} & 0.003 & 0.001 & 84.5 & \textbf{20} & \textbf{50} \\
\textbf{SS-C} & \textbf{0.001} & \textbf{0.001} & \textbf{0.003} & \textbf{0.004} & \textbf{0.004} & \textbf{0} & 247 & 101 & \textbf{50} \\
\textbf{SS-D} & \textbf{0} & \textbf{0.004} & \textbf{0.014} & \textbf{0.002} & 0.007 & 0.009 & 119.5 & 67 & \textbf{50} \\
\textbf{SS-E} & \textbf{0.001} & \textbf{0.001} & \textbf{0.012} & \textbf{0.004} & 0.008 & 0.002 & 208 & 54.5 & \textbf{50} \\
\textbf{SS-F} & \textbf{0} & 0.016 & 0.008 & \textbf{0} & 0.006 & 0.016 & 138 & 71 & \textbf{50} \\
\textbf{SS-G} & \textbf{0} & \textbf{0} & \textbf{0.023} & \textbf{0.003} & 0.006 & 0.004 & 131 & 69 & \textbf{50} \\
\textbf{SS-H} & \textbf{0} & \textbf{0} & \textbf{0} & \textbf{0} & \textbf{0} & \textbf{0} & 52 & \textbf{28} & \textbf{50} \\
\textbf{SS-I} & 0.008 & 0.018 & \textbf{0} & 0.008 & 0.018 & \textbf{0} & 48 & \textbf{30} & \textbf{50} \\
\textbf{SS-J} & \textbf{0} & \textbf{0} & \textbf{0.002} & \textbf{0.002} & \textbf{0.002} & \textbf{0} & 186 & \textbf{30} & \textbf{50} \\
\textbf{SS-K} & \textbf{0.001} & \textbf{0.001} & \textbf{0.003} & \textbf{0.001} & \textbf{0.002} & \textbf{0.001} & 209 & 140 & \textbf{50} \\
\textbf{SS-L} & \textbf{0.01} & \textbf{0.028} & \textbf{0.006} & \textbf{0.007} & \textbf{0.008} & \textbf{0.009} & 68.5 & \textbf{35} & \textbf{50} \\
\textbf{SS-M} & X & X & \textbf{0} & X & X & \textbf{0} & X & X & \textbf{50} \\
\textbf{SS-N} & X & X & \textbf{0.065} & X & X & \textbf{0.015} & X & X & \textbf{50} \\
\textbf{SS-O} & X & X & \textbf{3.01E-07} & X & X & \textbf{3.20E-06} & X & X & \textbf{50} \\ \midrule
\textbf{Win (\%)} & 73 & 67 & \textbf{93} & \textbf{67} & 33 & \textbf{67} & 0 & 33 & \textbf{80} \\ \bottomrule
\end{tabular}

\end{table*}

\noindent\textbf{RQ4: Can \flash be used to reduce the effort of multi-objective performance configuration optimization compared to ePAL?}\\
In the RQ4 section (right-hand side) of Table~\ref{table:multi-config}, the number of measurements required by methods, ePAL, and \flash are shown. Rows show different software systems and columns shows the number of measurements associated with each method. The numbers highlighted in bold mark methods that are statistically better than the other. For example in SS-K, \flash uses statistically fewer samples than (variants of) ePAL. 

From the table we observe:
\begin{itemize}
    \item \flash uses fewer samples than ePAL\_0.01. In 9 of 15 cases, ePAL\_0.01 is, at least, two times better than \flash.
    \item (Variants of) ePAL does not terminate for SS-M, SS-N, and SS-O even after ten hours of execution---a pragmatic choice. The reason for this can be seen in Table~\ref{fig:software_systems}: these software systems have more than 10 configuration options and the GPMs used by ePAL does not scale beyond that number. 
    \item The obvious feature of Table~\ref{table:multi-config} is that \flash used fewer measurements in 12 of 15 software systems.
\end{itemize}
\vskip 1ex
 \begin{myshadowbox}
         \flash requires fewer measurements than ePAL to approximate Pareto-optimal solutions. The number of evaluations used by \flash is less than (more careful) ePAL-0.01 for all the software systems and 12 of 15 software systems for (more careless) ePAL-0.3.
 \end{myshadowbox}

\begin{figure*}[htp]
\centering
\includegraphics[width=0.85\paperwidth, height=7cm]{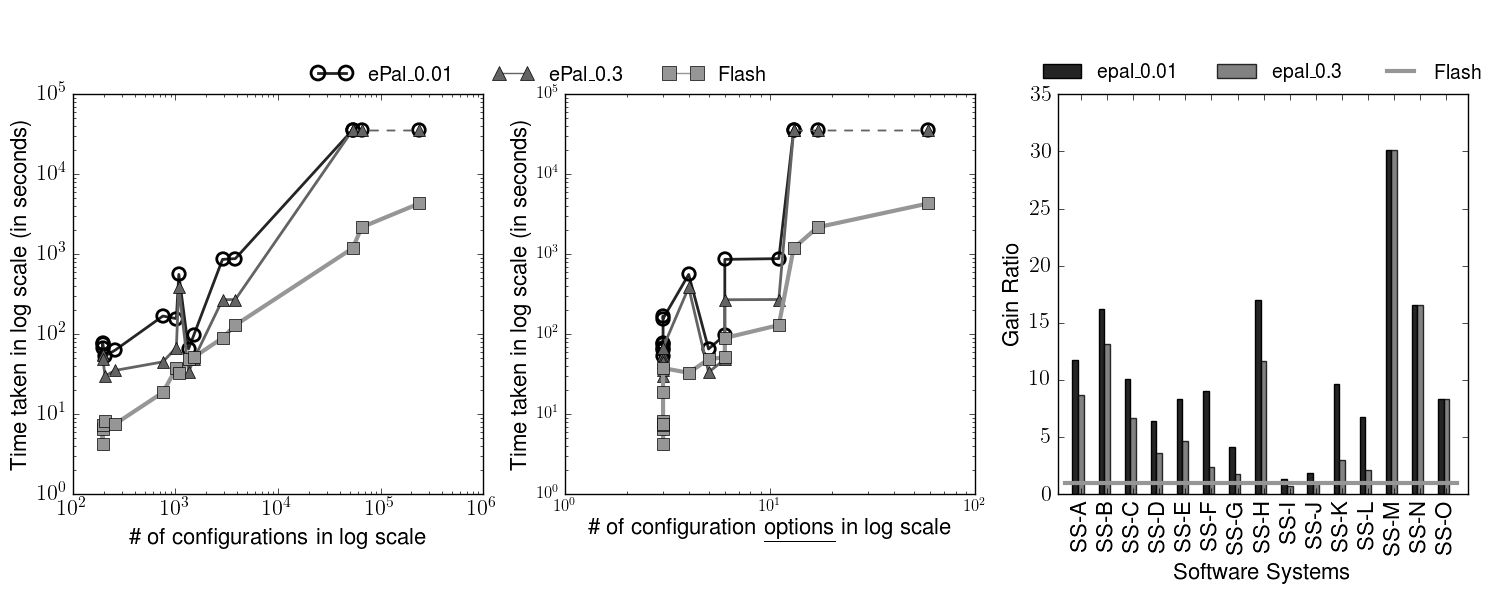}
\caption{The time required to find better solutions using ePAL and Flash (sum of 20 repeats). Note that the axis\textquotesingle s of the first two figures (left, and center) are in log scale. The time required for \flash compared to (variants of) ePAL is much lower (with an exception on 2 of 15 software systems). The dashed line in the figure (left and middle) represents cases where ePAL did not terminate within a reasonable time (10 hours). In the right-hand figure, we show the performance gain (wrt. to time) achieved by using \flash. All the bars above the dashed line (y=1) performs worse than \flash.}
\label{fig:multiconfig_time}
\end{figure*}

\noindent\textbf{RQ5:Does \flash save  time  for  multi-objective  performance configuration optimization compared to ePAL?} \\
Figure~\ref{fig:multiconfig_time} compares the run times of ePAL with \flash. Please note that we use the author\textquotesingle s version of ePAL in our experiments, which is implemented in Matlab. However, \flash was implemented in Python. Even though this may not be a fair comparison, for the sake of completeness, we report the run-times of the test. The sub-figure to the left shows how the run times vary with the number of configurations of the system. The x-axis represents the number of configurations (in log scale), while the y-axis represents the time taken to perform 20 repeats in seconds (in log scale), which means lower the better. The dotted lines in the figure, shows the cases where a method (in this case, ePAL) did not terminate. The sub-figure in the middle represents how the run-time varies with the number of configuration options. The x-axis represents the number of configuration options, and the y-axis represents the time taken for 20 repeats in seconds (in log scale), which means lower the better. The sub-figure to the right represent the performance gain achieved by \flash over (variants of) ePAL. The x-axis shows the software systems, and the Y-axis represents the gain ratio.   Any bar higher than the line  (y=1) represent cases where \flash is better than ePAL. 

From the figure, we observe:
\begin{itemize}
    \item From sub-figures left and middle, \flash is much faster than (variants of) ePAL except in 2 of 15 cases. 
    \item The run times of ePAL increase exponentially with the number of configurations and configuration options, similar to the trend reported in the literature.
    \item (Variants of) ePAL does not terminate for cases with large numbers of configurations and configuration options, whereas \flash always terminates an order of magnitude faster than ePAL. This effect is magnified in case of a scenarios with large configuration space.
\end{itemize}
Overall our results indicate:
\vskip 0.5ex
 \begin{myshadowbox}
         \flash saves time and is faster than (variants of) ePAL in 13 of 15 cases. Furthermore, \flash is an order of magnitudes faster than ePAL in 5 of 15 software systems. In other 2 out of 15 cases, the \flash\textquotesingle s runtimes are similar to (variants of) ePAL.
 \end{myshadowbox}

\begin{figure}
\centering
\includegraphics[height=6cm,width=7cm]{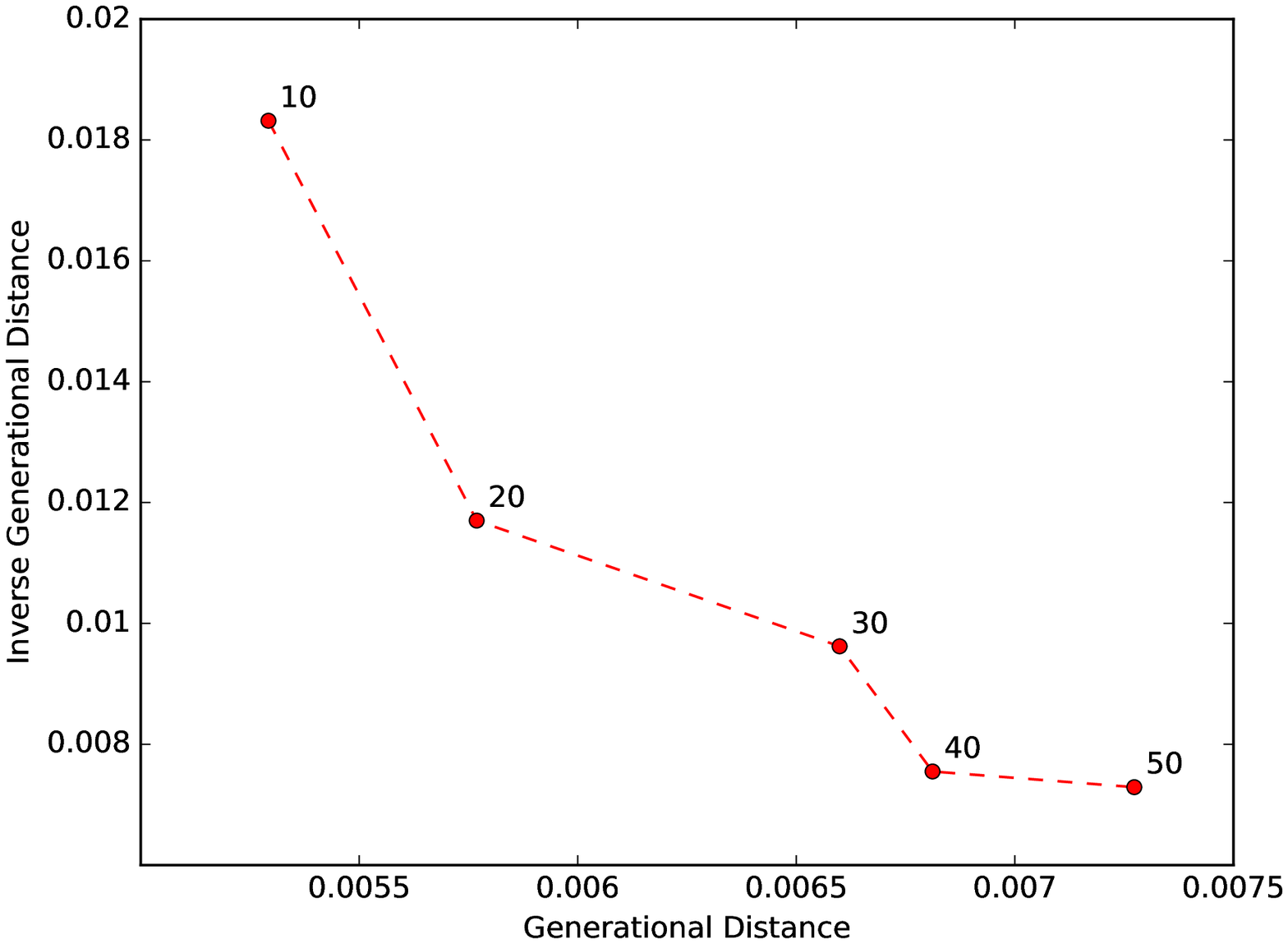}
\caption{The trade-off between the number of starting samples (exploration) and number of steps to converge (exploitation). The ideal point in these trade-off curves is (0,0), which mean the algorithm has perfect convergence ($GD=0$) and perfect diversity ($IGD=0$). The trade-off curve for multi-objective performance configuration optimization is shown with budget of 50 evaluations.}\label{fig:param_tuning_multiconfig}
\end{figure} 

\section{Discussion}

\subsection{Why CART is used as the surrogate model?}
Decision Trees are a very simple way to learn rules from a set of examples and can be viewed as a tool for the analysis of a large dataset. The reason why we chose CART is two-fold. Firstly, it is shown to be \textit{scalable} and there is a growing interest to find new ways to speed up the decision tree learning process~\cite{su2006fast}. 
Secondly, a decision tree can \textit{describe} with the tree structure the dependence between the decisions and the objectives, which is useful for induction and comprehensibility.
These are the primary reasons for choosing decision-trees to replace Gaussian Process as the surrogate model for \flash.

\subsection{What is the trade-off between the starting size and budget of {\sc \textbf{Flash}}?}\label{sec:parameter_tuning}

There are two main parameters of \flash which require being set very carefully. In our setting, the parameters are \textit{size} and \textit{budget}. The parameter \textit{size} controls the exploration capabilities of \flash whereas parameter \textit{budget} controls the exploitation capabilities. 
In Figure~\ref{fig:param_tuning_multiconfig}, we show the trade-off 
between generational distance and inverted generational distance 
by varying parameters \textit{size} and \textit{budget}. 
The markers in Figure~\ref{fig:param_tuning_multiconfig} are annotated with the starting size of \flash. The trade-off characterizes the relationship between two conflicting objectives, for example in Figure~\ref{fig:param_tuning_multiconfig}, point (10) achieves high convergence (low GD value) but low diversity (high IGD value). Note, that the curves are an aggregate of the trade-off curves for all the software systems. 
From the figure~\ref{fig:param_tuning_multiconfig} we observe that: The number of initial samples (\textit{size} in Figure~\ref{fig:flash_frame}) determines the diversity of the solution. With ten initial samples the algorithm converges (lowest GD values) but lowest diversity (high IGD values). However, with  50 initial samples (random sampling) \flash achieves highest diversity (low IGD values) but lowest convergence (high GD values). We choose the starting size of 30 because it achieves a good trade-off between convergence and diversity. These values were used in the experiments in section~\ref{sec:mo_results}.

\subsection{Can rules learned by CART  guide the search?}
Currently, \flash does not explicitly reflect on the Decision Tree to select the next sample (or configuration). But, rules learned by Decision Tree can be used to guide the process of search. Though we have not tested this approach, a Decision Tree can be used to learn about importance of various configuration options which can be then used to recursively prune the configuration space, similar to the approach of Oh et al.~\cite{oh2017finding}. We hypothesize that this would make \flash more scalable and be used to much larger models. We leave this for future work. 

\subsection{What are the shortcomings of \flash?}  
\flash  suffers from the following shortcomings.
\begin{itemize}[leftmargin=*]
    \item \textbf{Parallelization}:  \flash like all sequential model-based approaches completes evaluating a configuration before evaluating a new one. However, in practice, this feature can lead to really long runtimes. A possible extension might be to evaluate multiple configurations in parallel. This has not been considered in this version of \flash and is something we leave for the future work.
    
    \item  \textbf{Non-stationary}: \flash assumes that the benchmark or the load in the system is stationary. Hence, there is no inherent mechanism in \flash which would adapt itself based on the change in workload. This non-stationary nature of the problem is a significant assumption and currently not addressed in this paper. Addressing this aspect may include an ensemble-based approach where a new model is built at a specified time interval. The importance of the model is defined by a time-dependent weight decay of the model, that is, older the model, the lower its significance (weight).
    
    \item \textbf{Cost Sensitivity}: \flash also assumes that the cost of evaluating all the configurations are same. However, in practice, this is not true. For example, the wall clock time of running a specific benchmark on a software system with and without caching can be substantially different. In practice, stakeholders may demand to find a good configuration within a specified time limit (instead of the number of configurations measured). We leave this for future work.
    
    \item \textbf{Cold Start}: \flash randomly selects the initial configurations to evaluate, which can affect its effectiveness. One of the way to reduce the impact of randomness its to select the initial points based on domain knowledge or use transfer learning from similar software systems that have been optimized in the past, to select the initial configurations of \flash.
\end{itemize}

\section{Threats to Validity}
{\em Reliability} refers to the consistency of the results obtained
from the research.  For example,   how well can independent researchers reproduce the study? To increase external
reliability, we took care to either  define our
algorithms or use implementations from the public domain
(SciKitLearn)~\cite{scikit-learn}. All code used in this work are available
online~\footnote{\url{http://tiny.cc/flashrepo/}}.

{\em Validity} refers to the extent to which a piece of research
investigates what the researcher purports to investigate~\cite{SSA15}.
{\em Internal validity} concerns with whether the differences found in
the treatments can be ascribed to the treatments under study. 

For the case-studies relating to configuration control, we cannot measure all possible configurations in a reasonable time. Hence, we sampled only a few hundred configurations to compare the prediction to actual values. We are aware that this evaluation leaves room for outliers and that \textit{measurement bias} can cause false interpretations~\cite{me12d}. We also limit our attention to predicting PF for a given workload; we did not vary benchmarks.

\textit{Internal bias} originates from the stochastic nature of multi-objective optimization algorithms. The evolutionary process required many
random operations, same as the \flash was introduced in this article.
To mitigate these threats, we repeated our experiments for 20 runs
and reported the median of the indicators. We also employed
statistical tests to check the significance of the achieved results.

It is challenging to find the representatives sample test cases to covers all kinds of software systems. We just selected six most
common types of software system to discuss the \flash basing on
them. In the future, we also need to explore more types of SBSE
problems for other domains such as process planning, next release planning. We aimed to increase {\em external validity} by choosing case-studies from different domains. 

\section{Conclusion}

This article proposes a sequential model-based method called \flash, an approach for finding better configurations while minimizing the number of measurements. To the best of our knowledge, this is the first time a sequential model-based method is used to solve the problem of performance configuration optimization. \flash sequentially gains knowledge about the configuration space like traditional SMBO. \flash is different from the traditional SMBOs because of the choice of the surrogate model (CART) and the acquisition function (the stochastic Maximum-Mean {\em Bazza} function). We have demonstrated the effectiveness of \flash on single-objective and multi-objective problems using 30 scenarios from 7 software systems.

For a single-objective setting, we experimentally demonstrate that \flash can locate the better configuration of 30 different scenarios for seven software systems, accurately compared to the state-of-the-art approaches while removing the need for a holdout dataset, hence saving measurement costs. In 57\% of the scenarios, \flash can find the better configuration by using an order of magnitude fewer solutions than other state-of-the-art approaches.

For multi-objective setting, we show how \flash can overcome the shortcomings of traditional SMBO (ePAL) while being as effective as ePAL as well as being scalable to software systems with higher number (greater than 10) of configuration options (where ePAL does not terminate in a reasonable time-frame).


Regarding future work, the two directions for this research are i) test on different case studies and ii) further improve the scalability of \flash. To conclude, we urge the SE community to learn from communities which tackle similar problems. This article experiments with ideas from fields of machine learning, SBSE, and software analytics to create \flash, which is a fast, scalable and effective optimizer. We hope this article inspires other researchers to look further afield than their home discipline.

\section{Acknowledgement}
Apel\textquotesingle s work has been supported by the German Research Foundation (AP 206/6, AP 206/7, and AP 206/11).

\bibliographystyle{plain}


\begin{thebibliography}{10}

\bibitem{van2017automatic}
Dana Van~Aken, Andrew Pavlo, Geoffrey~J Gordon, and Bohan Zhang.
\newblock Automatic database management system tuning through large-scale
  machine learning.
\newblock In {\em International Conference on Management of Data}. ACM, 2017.

\bibitem{herodotou2011starfish}
Herodotos Herodotou, Harold Lim, Gang Luo, Nedyalko Borisov, Liang Dong,
  Fatma~Bilgen Cetin, and Shivnath Babu.
\newblock Starfish: A self-tuning system for big data analytics.
\newblock In {\em Conference on Innovative Data Systems Research}, 2011.

\bibitem{xu2015hey}
Tianyin Xu, Long Jin, Xuepeng Fan, Yuanyuan Zhou, Shankar Pasupathy, and Rukma
  Talwadker.
\newblock Hey, you have given me too many knobs!: understanding and dealing
  with over-designed configuration in system software.
\newblock In {\em Foundations of Software Engineering}, 2015.

\bibitem{zhu2017optimized}
Han Zhu, Junqi Jin, Chang Tan, Fei Pan, Yifan Zeng, Han Li, and Kun Gai.
\newblock Optimized cost per click in taobao display advertising.
\newblock {\em arXiv preprint}, 2017.

\bibitem{alipourfard2017cherrypick}
Omid Alipourfard, Hongqiang~Harry Liu, Jianshu Chen, Shivaram Venkataraman,
  Minlan Yu, and Ming Zhang.
\newblock Cherrypick: Adaptively unearthing the best cloud configurations for
  big data analytics.
\newblock In {\em Symposium on Networked Systems Design and Implementation},
  2017.

\bibitem{snoek2012practical}
Jasper Snoek, Hugo Larochelle, and Ryan~P Adams.
\newblock Practical bayesian optimization of machine learning algorithms.
\newblock In {\em Advances in neural information processing systems}, 2012.

\bibitem{brochu2010tutorial}
Eric Brochu, Vlad~M Cora, and Nando De~Freitas.
\newblock A tutorial on bayesian optimization of expensive cost functions, with
  application to active user modeling and hierarchical reinforcement learning.
\newblock {\em arXiv preprint}, 2010.

\bibitem{guo2013variability}
Jianmei Guo, Krzysztof Czarnecki, Sven Apel, Norbert Siegmund, and Andrzej
  Wasowski.
\newblock Variability-aware performance prediction: A statistical learning
  approach.
\newblock In {\em Automated Software Engineering}, 2013.

\bibitem{sarkar2015cost}
Atri Sarkar, Jianmei Guo, Norbert Siegmund, Sven Apel, and Krzysztof Czarnecki.
\newblock Cost-efficient sampling for performance prediction of configurable
  systems (t).
\newblock In {\em Automated Software Engineering}, 2015.

\bibitem{nair17}
Vivek Nair, Tim Menzies, Norbert Siegmund, and Sven Apel.
\newblock Faster discovery of faster system configurations with spectral
  learning.
\newblock {\em Automated Software Engineering}, 2017.

\bibitem{nair2017using}
Vivek Nair, Tim Menzies, Norbert Siegmund, and Sven Apel.
\newblock Using bad learners to find good configurations.
\newblock In {\em Foundations of Software Engineering}. ACM, 2017.

\bibitem{guo2017data}
Jianmei Guo, Dingyu Yang, Norbert Siegmund, Sven Apel, Atrisha Sarkar, Pavel
  Valov, Krzysztof Czarnecki, Andrzej Wasowski, and Huiqun Yu.
\newblock Data-efficient performance learning for configurable systems.
\newblock {\em Empirical Software Engineering}, pages 1--42, 2017.

\bibitem{zuluaga2016varepsilon}
Marcela Zuluaga, Andreas Krause, and Markus P{\"u}schel.
\newblock $\varepsilon$-pal: an active learning approach to the multi-objective
  optimization problem.
\newblock {\em Journal of Machine Learning Research}, 2016.

\bibitem{wang2016bayesian}
Ziyu Wang, Frank Hutter, Masrour Zoghi, David Matheson, and Nando de~Feitas.
\newblock Bayesian optimization in a billion dimensions via random embeddings.
\newblock {\em Journal of Artificial Intelligence Research}, 2016.

\bibitem{jamshidi2016uncertainty}
Pooyan Jamshidi and Giuliano Casale.
\newblock An uncertainty-aware approach to optimal configuration of stream
  processing systems.
\newblock In {\em Modeling, Analysis and Simulation of Computer and
  Telecommunication Systems}, 2016.

\bibitem{siegmund2012predicting}
Norbert Siegmund, Sergiy~S Kolesnikov, Christian K{\"a}stner, Sven Apel, Don
  Batory, Marko Rosenm{\"u}ller, and Gunter Saake.
\newblock Predicting performance via automated feature-interaction detection.
\newblock In {\em International Conference on Software Engineering}, 2012.

\bibitem{wang2013searching}
Tiantian Wang, Mark Harman, Yue Jia, and Jens Krinke.
\newblock Searching for better configurations: a rigorous approach to clone
  evaluation.
\newblock In {\em Foundations of Software Engineering}, 2013.

\bibitem{zuluaga2013active}
Marcela Zuluaga, Guillaume Sergent, Andreas Krause, and Markus P{\"u}schel.
\newblock Active learning for multi-objective optimization.
\newblock {\em International Conference of Machine Learning}, 2013.

\bibitem{chen2016sampling}
Jianfeng Chen, Vivek Nair, Rahul Krishna, and Tim Menzies.
\newblock Is sampling better than evolution for search-based software
  engineering?
\newblock {\em arXiv preprint}, 2016.

\bibitem{henard2015combining}
Christopher Henard, Mike Papadakis, Mark Harman, and Yves Le~Traon.
\newblock Combining multi-objective search and constraint solving for
  configuring large software product lines.
\newblock In {\em International Conference on Software Engineering}, 2015.

\bibitem{bergstra2013making}
James Bergstra, Daniel Yamins, and David Cox.
\newblock Making a science of model search: Hyperparameter optimization in
  hundreds of dimensions for vision architectures.
\newblock In {\em International Conference on Machine Learning}, 2013.

\bibitem{fu2016tuning}
Wei Fu, Tim Menzies, and Xipeng Shen.
\newblock Tuning for software analytics: Is it really necessary?
\newblock {\em Information and Software Technology}, 2016.

\bibitem{fufse17}
Wei Fu and Tim Menzies.
\newblock Easy over hard: A case study on deep learning.
\newblock In {\em Foundations of Software Engineering}. ACM, 2017.

\bibitem{fu2016differential}
Wei Fu, Vivek Nair, and Tim Menzies.
\newblock Why is differential evolution better than grid search for tuning
  defect predictors?
\newblock {\em arXiv preprint arXiv:1609.02613}, 2016.

\bibitem{tantithamthavorn2016automated}
Chakkrit Tantithamthavorn, Shane McIntosh, Ahmed~E Hassan, and Kenichi
  Matsumoto.
\newblock Automated parameter optimization of classification techniques for
  defect prediction models.
\newblock In {\em International Conference on Software Engineering}. IEEE,
  2016.

\bibitem{agrawal2016wrong}
Amritanshu Agrawal, Wei Fu, and Tim Menzies.
\newblock What is wrong with topic modeling?(and how to fix it using
  search-based se).
\newblock {\em arXiv preprint}, 2016.

\bibitem{venkataraman2016ernest}
Shivaram Venkataraman, Zongheng Yang, Michael~J Franklin, Benjamin Recht, and
  Ion Stoica.
\newblock Ernest: Efficient performance prediction for large-scale advanced
  analytics.
\newblock In {\em Symposium on Networked Systems Design and Implementation},
  2016.

\bibitem{yadwadkar2017selecting}
Neeraja~J. Yadwadkar, Bharath Hariharan, Joseph~E. Gonzalez, Burton Smith, and
  Randy~H. Katz.
\newblock Selecting the best vm across multiple public clouds: A data-driven
  performance modeling approach.
\newblock In {\em Symposium on Cloud Computing}. ACM, 2017.

\bibitem{Zhu:2017:BTP:3127479.3128605}
Yuqing Zhu, Jianxun Liu, Mengying Guo, Yungang Bao, Wenlong Ma, Zhuoyue Liu,
  Kunpeng Song, and Yingchun Yang.
\newblock Bestconfig: Tapping the performance potential of systems via
  automatic configuration tuning.
\newblock In {\em Symposium on Cloud Computing}. ACM, 2017.

\bibitem{dalibard2017boat}
Valentin Dalibard, Michael Schaarschmidt, and Eiko Yoneki.
\newblock Boat: Building auto-tuners with structured bayesian optimization.
\newblock In {\em Proceedings of the 26th International Conference on World
  Wide Web}. International World Wide Web Conferences Steering Committee, 2017.

\bibitem{biedermann2014hot}
Sebastian Biedermann, Stefan Katzenbeisser, and Jakub Szefer.
\newblock Hot-hardening: getting more out of your security settings.
\newblock In {\em Computer Security Applications Conference}. ACM, 2014.

\bibitem{biedermann2014leveraging}
Sebastian Biedermann, Stefan Katzenbeisser, and Jakub Szefer.
\newblock Leveraging virtual machine introspection for hot-hardening of
  arbitrary cloud-user applications.
\newblock In {\em HotCloud}, 2014.

\bibitem{drabik2003method}
John Drabik.
\newblock Method and apparatus for automatic configuration and management of a
  virtual private network, October~8 2003.
\newblock US Patent App. 10/460,518.

\bibitem{security1}
Hpe security research.
\newblock \url{http://files.asset.microfocus.com/4aa5-0858/en/4aa5-0858.pdf},
  2015.
\newblock [Online; accessed 10-Nov-2017].

\bibitem{security2}
Real-world access control.
\newblock
  \url{https://www.schneier.com/blog/archives/2009/09/real-world_acce.html},
  2009.
\newblock [Online; accessed 10-Nov-2017].

\bibitem{hill2017efficient}
Daniel~N Hill, Houssam Nassif, Yi~Liu, Anand Iyer, and SVN Vishwanathan.
\newblock An efficient bandit algorithm for realtime multivariate optimization.
\newblock In {\em SIGKDD International Conference on Knowledge Discovery and
  Data Mining}. ACM, 2017.

\bibitem{wang2016beyond}
Yue Wang, Dawei Yin, Luo Jie, Pengyuan Wang, Makoto Yamada, Yi~Chang, and
  Qiaozhu Mei.
\newblock Beyond ranking: Optimizing whole-page presentation.
\newblock In {\em International Conference on Web Search and Data Mining}. ACM,
  2016.

\bibitem{Siegmund2015}
Norbert Siegmund, Alexander Grebhahn, Sven Apel, and Christian K\"{a}stner.
\newblock Performance-influence models for highly configurable systems.
\newblock In {\em Foundations of Software Engineering}. ACM, 2015.

\bibitem{shen2006fast}
Yirong Shen, Matthias Seeger, and Andrew~Y Ng.
\newblock Fast gaussian process regression using kd-trees.
\newblock In {\em Advances in neural information processing systems}, 2006.

\bibitem{krall2015gale}
Joseph Krall, Tim Menzies, and Misty Davies.
\newblock Gale: Geometric active learning for search-based software
  engineering.
\newblock {\em IEEE Transactions on Software Engineering}, 2015.

\bibitem{breiman1984classification}
Leo Breiman, Jerome~H Friedman, Richard~A Olshen, and Charles~J Stone.
\newblock {\em Classification and regression trees}.
\newblock Wadsworth \& Brooks/Cole Advanced Books \& Software, 1984.

\bibitem{zhang2007moea}
Qingfu Zhang and Hui Li.
\newblock Moea/d: A multiobjective evolutionary algorithm based on
  decomposition.
\newblock {\em IEEE Transactions on Evolutionary Computation}, 2007.

\bibitem{hoffman2014modular}
Matthew~W Hoffman and Bobak Shahriari.
\newblock Modular mechanisms for bayesian optimization.
\newblock In {\em NIPS Workshop on Bayesian Optimization}, pages 1--5.
  Citeseer, 2014.

\bibitem{bergstra2011algorithms}
James~S Bergstra, R{\'e}mi Bardenet, Yoshua Bengio, and Bal{\'a}zs K{\'e}gl.
\newblock Algorithms for hyper-parameter optimization.
\newblock In {\em Advances in Neural Information Processing Systems}, 2011.

\bibitem{harman2009search}
Mark Harman, S~Afshin Mansouri, and Yuanyuan Zhang.
\newblock Search based software engineering: A comprehensive analysis and
  review of trends techniques and applications.
\newblock {\em Department of Computer Science, King’s College London, Tech.
  Rep. TR-09-03}, 2009.

\bibitem{sarro2016multi}
Federica Sarro, Alessio Petrozziello, and Mark Harman.
\newblock Multi-objective software effort estimation.
\newblock In {\em Proceedings of the 38th International Conference on Software
  Engineering}, pages 619--630. ACM, 2016.

\bibitem{chen2017beyond}
Jianfeng Chen, Vivek Nair, and Tim Menzies.
\newblock Beyond evolutionary algorithms for search-based software engineering.
\newblock {\em arXiv preprint}, 2017.

\bibitem{golovin2017google}
Daniel Golovin, Benjamin Solnik, Subhodeep Moitra, Greg Kochanski, John Karro,
  and D~Sculley.
\newblock Google vizier: A service for black-box optimization.
\newblock In {\em International Conference on Knowledge Discovery and Data
  Mining}. ACM, 2017.

\bibitem{shahriari2016taking}
Bobak Shahriari, Kevin Swersky, Ziyu Wang, Ryan~P Adams, and Nando De~Freitas.
\newblock Taking the human out of the loop: A review of bayesian optimization.
\newblock {\em Proceedings of the IEEE}, 104(1):148--175, 2016.

\bibitem{rasmussen2004gaussian}
Carl~Edward Rasmussen.
\newblock Gaussian processes in machine learning.
\newblock In {\em Advanced lectures on machine learning}, pages 63--71.
  Springer, 2004.

\bibitem{su2006fast}
Jiang Su and Harry Zhang.
\newblock A fast decision tree learning algorithm.
\newblock In {\em AAAI Conference on Artificial Intelligence}, 2006.

\bibitem{domingos2000mining}
Pedro Domingos and Geoff Hulten.
\newblock Mining high-speed data streams.
\newblock In {\em Proceedings of the sixth ACM SIGKDD international conference
  on Knowledge discovery and data mining}, pages 71--80. ACM, 2000.

\bibitem{deb2002fast}
Kalyanmoy Deb, Amrit Pratap, Sameer Agarwal, and TAMT Meyarivan.
\newblock A fast and elitist multiobjective genetic algorithm: Nsga-ii.
\newblock {\em IEEE transactions on evolutionary computation}, 2002.

\bibitem{oh2017finding}
Jeho Oh, Don Batory, Margaret Myers, and Norbert Siegmund.
\newblock Finding near-optimal configurations in product lines by random
  sampling.
\newblock In {\em Foundations of Software Engineering}, 2017.

\bibitem{wang2016practical}
Shuai Wang, Shaukat Ali, Tao Yue, Yan Li, and Marius Liaaen.
\newblock A practical guide to select quality indicators for assessing
  pareto-based search algorithms in search-based software engineering.
\newblock In {\em International Conference on Software Engineering}, 2016.

\bibitem{van1999multiobjective}
David~Allen Van~Veldhuizen.
\newblock Multiobjective evolutionary algorithms: Classifications, analyses,
  and new innovations.
\newblock Technical report, 1999.

\bibitem{coello2004study}
Carlos A~Coello Coello and Margarita~Reyes Sierra.
\newblock A study of the parallelization of a coevolutionary multi-objective
  evolutionary algorithm.
\newblock In {\em Mexican International Conference on Artificial Intelligence},
  2004.

\bibitem{mittas13}
N.~Mittas and L.~Angelis.
\newblock Ranking and clustering software cost estimation models through a
  multiple comparisons algorithm.
\newblock {\em IEEE Transactions on Software Engineering}, 39, 2013.

\bibitem{efron93}
B.~Efron and R.~J. Tibshirani.
\newblock {\em {An Introduction to the Bootstrap}}.
\newblock CRC, 1993.

\bibitem{shepperd12a}
Martin~J. Shepperd and Steven~G. MacDonell.
\newblock Evaluating prediction systems in software project estimation.
\newblock {\em Information {\&} Software Technology}, 54(8):820--827, 2012.

\bibitem{kampenes07}
Vigdis~By Kampenes, Tore Dyb{\aa}, Jo~Erskine Hannay, and Dag I.~K. Sj{\o}berg.
\newblock A systematic review of effect size in software engineering
  experiments.
\newblock {\em Information {\&} Software Technology}, 49(11-12):1073--1086,
  2007.

\bibitem{Kocaguneli2013:ep}
Ekrem Kocaguneli, Thomas Zimmermann, Christian Bird, Nachiappan Nagappan, and
  Tim Menzies.
\newblock {Distributed development considered harmful?}
\newblock In {\em Proceedings - International Conference on Software
  Engineering}, pages 882--890, 2013.

\bibitem{scikit-learn}
Fabian Pedregosa, Ga{\"e}l Varoquaux, Alexandre Gramfort, Vincent Michel,
  Bertrand Thirion, Olivier Grisel, Mathieu Blondel, Peter Prettenhofer, Ron
  Weiss, Vincent Dubourg, et~al.
\newblock Scikit-learn: Machine learning in python.
\newblock {\em Journal of Machine Learning Research}, 2011.

\bibitem{SSA15}
Janet Siegmund, Norbert Siegmund, and Sven Apel.
\newblock Views on internal and external validity in empirical software
  engineering.
\newblock In {\em International Conference on Software Engineering}. IEEE,
  2015.

\bibitem{me12d}
Tim Menzies, Andrew Butcher, David Cok, Andrian Marcus, Lucas Layman, Forrest
  Shull, Burak Turhan, and Thomas Zimmermann.
\newblock Local versus global lessons for defect prediction and effort
  estimation.
\newblock {\em IEEE Transactions on Software Engineering}, 2013.

\end{thebibliography}



\end{document}